\documentclass[aps,prd,11pt,showpacs,superscriptaddress,amssymb,amsmath,nofootinbib]{revtex4-1}

\usepackage{hhline}
\usepackage{color}
\usepackage{epsfig}
\usepackage{graphicx}
\usepackage{float}%
\usepackage{epstopdf}
\usepackage{hyperref}
\usepackage[usenames,dvipsnames]{xcolor}
\usepackage{multirow,booktabs}
\usepackage[normalem]{ulem}
\usepackage[d]{esvect}
\usepackage[misc,geometry]{ifsym}
\usepackage{harpoon}
\usepackage{subcaption}
\usepackage{graphicx}
\usepackage{cleveref}

\begin{document}

	\title{ Gravitational-lensing-modulated leptonic CP violation}
	\author{Ze-Wen Li}
	\author{Shu-Jun Rong}
    \author{Ya-Ru Wang}
	\affiliation{College of Physics and Electronic Information Engineering, Guilin University of Technology, Guilin, Guangxi 541004, China}

	\begin{abstract}
	The discovery of high-energy astrophysical neutrinos has recently prompted interest in the role of gravity in neutrino flavor conversion.
In this work, we investigate the gravity-modulated CP violation (CPV) in neutrino oscillations in
 the Reissner--Nordstr\"om and Hayward spacetimes. Employing analytical and numerical methods, we reveal the interplay among the CPV, the properties of neutrinos, and the space-time.
 The shapes of the flavor-oscillation curves  show that  CPV signals encode both space-time geometry and mass ordering, and that the imprints of $m_l$ are intrinsically correlated with the singular versus nonsingular structure of the background metric.
In particular, our finding demonstrate that the CPV observables exceed the limitation that, in asymptotically flat space-time, gravity neither damps nor amplifies neutrino-oscillation amplitudes.
		
	\end{abstract}
	

	\maketitle

\section{Introduction}

The charge-conjugation and parity  violation (CPV) has long been regarded as one of the central issues in high-energy physics. Its importance is closely connected with the observed matter--antimatter asymmetry of the Universe, whose dynamical generation requires C and CP violation, baryon-number violation, and departure from thermal equilibrium as the Sakharov conditions \cite{Sakharov:1967dj,Hambye:2012zz}. Although CPV was first established experimentally in the neutral kaon system \cite{Christenson:1964fg} and is described in the Standard Model through the complex phase of the Cabibbo--Kobayashi--Maskawa (CKM) matrix \cite{Nir:1992uv,Branco:2011zb,Kobayashi:1973fv,Branco:1999fs}, the CPV processes within the Standard Model seem unable to produce the observed baryon asymmetry \cite{Nir:1992uv}. This limitation motivates the exploration of CPV mechanisms beyond the Standard Model.

Neutrino oscillations provide a natural framework for probing leptonic CPV. In the three-flavor framework, flavor neutrino states are superpositions of massive neutrino states through the leptonic mixing matrix \cite{Bilenky:2018hbz}. As a result, neutrino oscillation probabilities depend not only on the propagation phases of the mass eigenstates, but also on the complex structure of the mixing matrix, making oscillations a natural probe of leptonic CPV \cite{Branco:2011zb,Sengupta:2026qig,Camara:2026ijm,Dutta:2026vhl,Pandit:2026fyp,Das:2026rel,Granelli:2026qss,Ignoti:2025rxr,Chundawat:2026lcm,Yadav:2026lsx,Nambiar:2026zhs}. This possibility is also of cosmological interest: in leptogenesis scenarios, a lepton asymmetry generated by lepton-number-violating processes can be partially converted into a baryon asymmetry through sphaleron processes \cite{Hambye:2012zz}. 

The difference between neutrino and antineutrino transition probabilities defines the CP asymmetry,
\begin{equation}\label{eq:cpv}
A_{\alpha\beta}^{\rm CP}
\equiv
P(\nu_\alpha\rightarrow\nu_\beta)
-
P(\bar{\nu}_\alpha\rightarrow\bar{\nu}_\beta),
\end{equation}
which is one of the key observables for testing CPV in the lepton sector. However, a measured CP asymmetry does not necessarily reflect only the intrinsic Dirac phase; it can also be modified by matter effects, nonstandard interactions, non-unitary mixing, or other changes in neutrino propagation \cite{Capolupo:2018hrp,Parveen:2024cff,Gandhi:2015xza,Dutta:2016vcc,Giri:2025hdw,Bernabeu:2018twl,Cherchiglia:2023ojf,Alonso-Alvarez:2024wnh,Yu:2024nkc}.

Gravity can provide another possible modulation mechanism for neutrino oscillation observables in astrophysical environments. Around compact objects, spacetime curvature modifies the propagation phase accumulated by each neutrino mass eigenstate \cite{Wudka:1991tg,Ahluwalia:1996ev,Fornengo:1996ef,Cardall:1996cd,Piriz:1996mu,Crocker:2003cw,Lambiase:2005gt,Ren:2010yf,
Sreekumar:2026glw,Wang:2026vgc,Gregoris:2025tpx,Wang:2024tfh,Lambiase:2023pxd,Buoninfante:2023qbk,Barick:2023yrs,Ettefaghi:2022nsq,Swami:2022xet,Sadeghi:2021zaw,Luciano:2021gdp,Mastrototaro:2021kmw,Mandal:2021dxk,
Boshkayev:2020igc,Dvornikov:2019jkc,Buoninfante:2019der,Zhang:2017ink,Yang:2017asl,Zhang:2016deq,Chakraborty:2015vla,Visinelli:2014xsa,Geralico:2012zt,Godunov:2009ce,Ren:2009zz}. When gravitational lensing (GL) is present, different propagation paths can interfere, further modifying the oscillation probability\cite{Fornengo:1996ef,Crocker:2003cw} and hence the CP asymmetry. Previous studies have shown that gravitationally induced phase corrections may encode information on the mass ordering and, in some cases, on the absolute neutrino mass scale \cite{Chakrabarty:2021bpr,Shi:2025plr,Chakrabarty:2023kld,Alloqulov:2024sns,Shi:2024flw}. This makes gravitational lensing a useful theoretical setting for examining the interplay among neutrino properties, CP asymmetry, and spacetime geometry.

Most existing studies on gravity-modified neutrino oscillations have focused on weak-field approximations or on a single gravitational background\cite{Shi:2025plr,Shi:2025ywa}. The role of distinct classes of spherically symmetric compact-object geometries, including singular and nonsingular configurations, and the impacts of field strength on modulating CPV observables  remain unexplored. We consider that such geometry-dependent and field strength effects are important because different metric functions modify the deflection angle, impact parameter, and accumulated phase in different ways, thereby changing the amplitude, period, phase shift, and even the sign of the CP asymmetry.

Gravitational-lensing-induced flavour interferometry in the Reissner--Nordstr\"om background has been studied
in a two-flavour framework under the weak-field approximation, where the charge was shown to shift the interference pattern~\cite{Alexandre:2025qip}. In this work, we investigate gravitational-lensing-modulated CP asymmetries in three-flavour neutrino oscillations, including the Dirac CP phase, and compare the results in the  Reissner--Nordstr\"om (RN) and Hayward~\cite{Hayward:2005gi} spacetimes. The Reissner--Nordstr\"om geometry represents a singular charged space-time with an irremovable central singularity \cite{Marsh:2007ib,Chesler:2019tco}, while the Hayward (HA) metric provides a nonsingular alternative in which the internal singularity is replaced by a regular center \cite{Hayward:2005gi}. At small radius, the HA core exhibits an effective cosmological-constant behavior characterized by the length scale $l$ \cite{Hayward:2005gi}. We derive the corresponding oscillation phases and impact parameters, and analyze the dependence of $A_{e\mu}^{\rm CP}$ on the mass ordering, the lightest neutrino mass, and the characteristic parameters of each metric. By comparing weak-field approximations with strong-field numerical calculations, we show how gravitational parameters can amplify, suppress, shift, or reshape CPV oscillation patterns.

The paper is organized as follows. In \cref{sec:Theo}, the general theoretical framework addressing neutrino oscillations in curved space-time is presented. In \cref{sec:lensing}, the analytical expressions of the oscillation phases based on the chosen metrics are derived.
In \cref{sec:Nume}, numerical results are shown with illustrative space-time parameters. In \cref{sec:Conc}, the conclusion is given. Throughout the paper, the units  $G=\hbar=c=1$ are employed.

\section{Theoretical framework}	
\label{sec:Theo}
\subsection{Neutrino oscillation in flat space-time}
\label{subsec:neuosc}
Neutrino oscillations\cite{Super-Kamiokande:1998kpq,SNO:2002tuh,DayaBay:2012fng} reveal that neutrinos have masses and a flavor state is a superposition of the massive states, namely
	\begin{equation}
		|\nu_{\alpha}\rangle = \sum_{i = 1}^{3}U_{\alpha i}^{\ast}|\nu_{i}\rangle,
		\label{eq:nu_i_to_alpha}
	\end{equation}
	with \(U_{\alpha i}\) being the $3\times 3$  Pontecorvo-Maki-Nakagawa-Sakata (PMNS) leptonic mixing matrix\cite{Pontecorvo:1957qd,
Maki:1962mu,Pontecorvo:1967fh}, i.e.,
\begin{equation}\label{eq:PMNS}
U=
\begin{pmatrix}
c_{12}c_{13}
& s_{12}c_{13}
& s_{13}e^{-i\delta_{\mathrm{CP}}}
\\
-s_{12}c_{23}-c_{12}s_{13}s_{23}e^{i\delta_{\mathrm{CP}}}
& c_{12}c_{23}-s_{12}s_{13}s_{23}e^{i\delta_{\mathrm{CP}}}
& c_{13}s_{23}
\\
s_{12}s_{23}-c_{12}s_{13}c_{23}e^{i\delta_{\mathrm{CP}}}
& -c_{12}s_{23}-s_{12}s_{13}c_{23}e^{i\delta_{\mathrm{CP}}}
& c_{13}c_{23}
\end{pmatrix},
\end{equation}
where $c_{ij}=\cos\theta_{ij}$ and $s_{ij}=\sin\theta_{ij}$
$(i,j=1,2,3)$.

For a neutrino produced at \(\left( t_{S},x_{S} \right)\ \) and detected at \(\left( t_{D},x_{D} \right)\), the propagation results in a phase, i.e.,
\begin{equation}
		\left| \nu_{i}\left( t_{D},x_{D} \right)\rangle = \exp\left( - i\Phi_{i} \right) \right|\nu_{i}\left( t_{S},x_{S} \right)\rangle.
		\label{eq:plane_wave}
	\end{equation}
The phase differences between the massive states lead to the flavor oscillation expressed by the probability
\begin{equation}
		\label{eq:prob_alpha_to_beta}
		P_{\alpha\beta} = \mid \langle\nu_{\beta}|\nu_{\alpha}\left( t_{D},x_{D} \right)\rangle \mid^{2}= \sum_{i,j}^{}U_{\beta i}U_{\beta j}^{\ast}U_{\alpha j}U_{\beta i}^{\ast}\exp\left[-i\left( \Phi_{i} - \Phi_{j} \right) \right].
	\end{equation}
In flat space-time, the phase differences, dependent on the squared mass difference \(\ \Delta m_{ij}^{2} = m_{i}^{2} - m_{j}^{2}\), are written as follow
	\begin{equation}
		\label{eq:flat_phase_diff}
		\Delta\Phi_{ij} \equiv \Phi_{i} - \Phi_{j} \simeq \frac{\Delta m_{ij}^{2}}{2E_{0}}\left|x_{D} - x_{S} \right|,
	\end{equation}
	with \(E_{0}\)  the average energy of neutrinos.

\subsection{Neutrino oscillation in curved space-time}
In curved spacetime,  the phase \(\Phi_{k}\) should be generalised to a covariant form \cite{Stodolsky:1978ks}
	\begin{equation}
		\label{eq:curved_phase}
		\Phi_{k} = \int_{S}^{D}p_{\mu}^{\left( k \right)}dx^{\mu},
	\end{equation}
	where
	\begin{equation}
		\label{eq:p_mu}
		{\ p}_{\mu}^{\left( k \right)} = m_{k}g_{\mu\nu}\frac{dx^{\nu}}{ds}
	\end{equation}
	is the canonical conjugate momentum to the coordinates \(x^{\mu}\),  $m_k$ satisfies the mass-shell equation
	\begin{equation}
		\label{eq:mass_shell}
		-m_k^2=g^{\mu\nu}p_\mu^{(k)}p_\nu^{(k)}.
	\end{equation}

For a spherically symmetric metric, its line element can be expressed in the following form
\begin{equation}\label{eq1}
ds^2 = -A \, dt^2 + B \, dr^2 + C \, d\theta^2 + D \, d\varphi^2,
\end{equation}
with
\begin{equation}\label{eq2}
\begin{aligned}
g_{tt} &= -A, \quad g^{tt} = -1/A,\\
g_{rr} &= B, \quad g^{rr} = 1/B,\\
g_{\theta\theta} &= C, \quad g^{\theta\theta} = 1/C,\\
g_{\varphi\varphi} &= D, \quad g^{\varphi\varphi} = 1/D,\\
C&= r^{2},\quad D= r^{2}\sin^{2}\theta.
\end{aligned}
\end{equation}

We denote the  momentum as
\begin{equation}\label{eq3}
\begin{aligned}
p_{t}^{(k)} &= m_k g_{tt} \frac{dt}{ds} \equiv -E_k,\\
p_{r}^{(k)} &= m_k g_{rr} \frac{dr}{ds} \equiv p_k,\\
p_{\varphi}^{(k)} &= m_k g_{\varphi\varphi} \frac{d\varphi}{ds} \equiv J_k,
\end{aligned}
\end{equation}
The mass-shell relation can be expressed as
\begin{equation}\label{eq4}
-m_k^2 = g^{tt} p_t^2 + g^{rr} p_r^2 + g^{\varphi\varphi}p_{\varphi}^{2}.
\end{equation}
For the non-radial propagation of neutrinos, its phase can be written as
\begin{equation}\label{eq5}
\Phi_k = \int_S^D \left[ - E_k\left(\frac{dt}{dr}\right)_0 + p_k + J_k\left(\frac{d\varphi}{dr}\right)_0 \right] dr,
\end{equation}
where the subscript '0' represents a light-ray differential\cite{Fornengo:1996ef}.

Without losing generality, we consider the motion on the $\theta=\pi/2$ plane. It follows from Eq.\eqref{eq3} that
\begin{equation}\label{eq6}
\left( \frac{d t}{d r} \right)_{0} = \frac{E_{0}}{p_{0}} \frac{B}{A},\quad
\left( \frac{d \varphi}{d r} \right)_{0} = \frac{J_{0}}{p_{0}} \frac{B}{D},
\end{equation}
where $E_{0}$, $J_{0}$, and $p_{0}$ are quantities of a massless particle.

The conserved angular momentum reads
\begin{equation}\label{eq7}
J_{k} = E_{k} b v_{k}^{(\infty)},
\end{equation}
where $b$ is the impact parameter.

The velocity \(v_{k}\) at infinity is
\begin{equation}\label{eq8}
v_{k}^{(\infty)} = \frac{\sqrt{E_{k}^{2} - m_{k}^{2}}}{E_{k}}
\simeq 1 - \frac{m_{k}^{2}}{2E_{k}^{2}}.
\end{equation}
Since the mass $m_k$ of the neutrino is extremely small, the angular momentum can be approximately calculated as
\begin{equation}\label{eq9}
J_{k} \simeq E_{k} b \left( 1 - \frac{m_{k}^{2}}{2E_{k}^{2}} \right).
\end{equation}
For massless particles
\begin{equation}\label{eq10}
J_{0} = E_{0} b.
\end{equation}
Substituting Eqs.\eqref{eq2} \eqref{eq3} \eqref{eq9}, and \eqref{eq10} into the mass-shell relation, we  obtain
\begin{equation}\label{eq11}
\begin{split}
p_k &= \pm E_k
\sqrt{\frac{B}{A}-\frac{Bb^2}{D}
-B\left(1-\frac{b^2}{D}\right)\frac{m_k^2}{E_k^2}},\\
p_0 &= \pm E_0
\sqrt{\frac{B}{A}-\frac{Bb^2}{D}}.
\end{split}
\end{equation}
After substituting all the above equations into Eq.\eqref{eq5}, the phase equation for non-radial propagation of neutrinos can be obtained as follow
\begin{equation}\label{eq12}
\Phi_{k}=\pm\frac{m_{k}^{2}}{2E_{0}}\int_{S}^{D}\sqrt{AB}\left(1-\frac{b^{2}A}{D}\right)^{-\frac{1}{2}}dr.
\end{equation}
This equation is applicable to all curved space-times described by spherically symmetric metrics.

\begin{figure}[hb]
		\centering
		\includegraphics[width=0.5\linewidth]{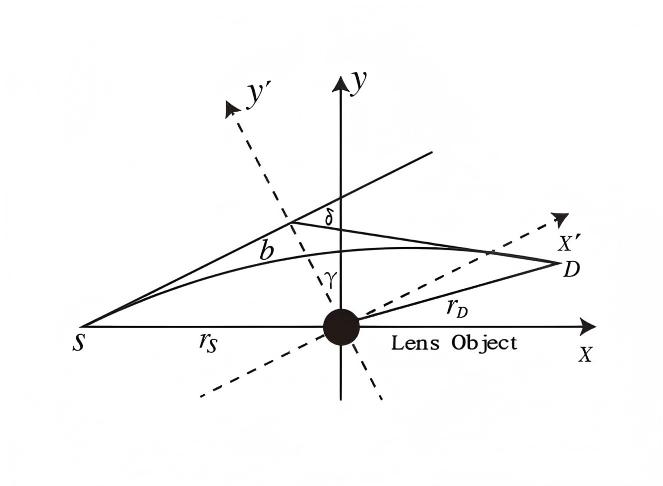}
		\caption{Schematic neutrino propagation impacted by gravitational lensing. $S$ is the source, $D$ is the detector. $b$ is the impact factor, $\delta$ is the deflection angle and $\gamma$ marks the misalignment of the coordinates $(x,y)$ and $(x',y')$.
			For the original construction of the plot, see \cite{Swami:2020qdi} and the adapted versions see \cite{Chakrabarty:2021bpr}.}
		\label{fig:lens}
	\end{figure}

We consider the case that neutrinos are produced at source S, propagate outward, and orbit around a massive celestial body, see Fig.\ref{fig:lens}. The motion passes through the closest point at $r = r_{0}$ and eventually reaches the detection point D.
Then the phase is calculated by the following equation
\begin{equation}\label{eq13}
\Phi_{k}\left(r_{S} \rightarrow r_{0} \rightarrow r_{D}\right) = \frac{m_{k}^{2}}{2E_{0}} \int_{r_{0}}^{r_{S}} \sqrt{\frac{AB}{1 - \frac{b^{2}A}{r^{2}}}} dr + \frac{m_{k}^{2}}{2E_{0}} \int_{r_{0}}^{r_{D}} \sqrt{\frac{AB}{1 - \frac{b^{2}A}{r^{2}}}} dr.
\end{equation}
Note that the shortest distance $r_{0}$  is determined by the equation
\begin{equation}\label{eq14}
\left(\frac{dr}{d\phi}\right)_{0} = \frac{p_0(r_0)D}{J_0 B} = 0.
\end{equation}
After substituting Eqs.\eqref{eq11} and \eqref{eq10} into \eqref{eq14}, it can be derived that
\begin{equation}\label{eq15}
B\left(D-A b^2\right)=0 .
\end{equation}
Based on the above phase, the probability of neutrino flavor oscillation impacted by GL reads
\begin{equation}\label{eq49}
P_{\alpha\beta} = \left| \langle \nu_\beta^D | \nu_\alpha^S \rangle \right|^2 = |N|^2 \sum_{i,j} U_{\beta i} U_{\beta j}^* U_{\alpha j} U_{\alpha i}^* \sum_{p,q} \exp\left(-i\Delta \Phi_{ij}^{pq}\right),
\end{equation}
where
\begin{equation}\label{eq50}
|N|^2 = \left( \sum_{i} |U_{\alpha i}|^2 \sum_{p,q} \exp\left(-i\Delta \Phi_{ii}^{pq}\right) \right)^{-1}
\end{equation}
is the normalization constant,
p and q represent different propagation paths. In this paper, we consider the case that each massive  neutrino has 2 non-radial paths corresponding to 2 impact parameters.
 For anti-neutrinos, oscillation probabilities can be obtained by taking the complex conjugate of the PMNS matrix.
Then the CPV is calculated according to the Eq.\eqref{eq:cpv}.

\section{Lensing by singular and nonsingular gravities }\label{sec:lensing}
To show the lensing effects on the CPV under spherically symmetric metrics, we consider both general relativity (GR)  and the gravity without singularities.  We choose the RN metric for the former and  the Hayward for the latter.
According to the general methodology in the previous section, we derive the analytical expressions for the oscillations phases and the corresponding impact parameters.

\subsection{Reissner--Nordstr\"om metric}
\subsubsection{Oscillation Phase}
The line element of the RN metric reads
\begin{equation}\label{eq40}
ds^2 = -\left(1 - \frac{2M}{r} + \frac{Q^2}{r^2}\right)dt^2 + \left(1 - \frac{2M}{r} + \frac{Q^2}{r^2}\right)^{-1}dr^2 + r^2 d\theta^2 + r^2 \sin^2\theta \, d\phi^2.
\end{equation}
Substitute Eqs.\eqref{eq40} and \eqref{eq2}  into Eq.\eqref{eq13}, we get
\begin{equation}\label{eq41}
\begin{split}
\Phi_k(r_S \to r_0 \to r_D) = \frac{m_k^2}{2E_0} \int_{r_0}^{r_S} \frac{1}{\sqrt{1 - \frac{b^2}{r^2}}} \left( 1 + \frac{2b^2 M}{r(r^2 - b^2)} - \frac{b^2 Q^2}{r^2(r^2 - b^2)} \right)^{-1/2} dr\\
+ \frac{m_k^2}{2E_0} \int_{r_0}^{r_D} \frac{1}{\sqrt{1 - \frac{b^2}{r^2}}} \left( 1 + \frac{2b^2 M}{r(r^2 - b^2)} - \frac{b^2 Q^2}{r^2(r^2 - b^2)} \right)^{-1/2} dr.
\end{split}
\end{equation}
Taking $ \frac{2b^2 M}{r(r^2 - b^2)} - \frac{b^2 Q^2}{r^2(r^2 - b^2)}$ as a small quantity,
Eq.\eqref{eq41} can be approximated as
\begin{equation}\label{eq42}
\begin{split}
\Phi_k(r_S\to r_0\to r_D)
&=\frac{m_k^2}{2E_0}\int_{r_0}^{r_S}
\frac{1}{\sqrt{1-\frac{b^2}{r^2}}}
\left(
1-\frac{b^2M}{r(r^2-b^2)}
+\frac{b^2Q^2}{2r^2(r^2-b^2)}
\right)\,dr\\
&\quad+\frac{m_k^2}{2E_0}\int_{r_0}^{r_D}
\frac{1}{\sqrt{1-\frac{b^2}{r^2}}}
\left(
1-\frac{b^2M}{r(r^2-b^2)}
+\frac{b^2Q^2}{2r^2(r^2-b^2)}
\right)\,dr.
\end{split}
\end{equation}
After integrating Eq.\eqref{eq42}, we obtain
\begin{equation}\label{eq43}
\begin{aligned}
\Phi_k &= \frac{m_k^2}{2E_0} \left. \left( \sqrt{r^2 - b^2} + \frac{M r}{\sqrt{r^2 - b^2}} - \frac{Q^2r \sqrt{1 - \frac{b^2}{r^2}}\text{Hypergeometric2F1}\left[ -\frac{1}{2},1, \frac{1}{2},  1 - \frac{r^2}{b^2} \right]}{2 (-b^2 + r^2)} \right) \right|_{r_0}^{r_S} \\
&\quad + \frac{m_k^2}{2E_0} \left. \left( \sqrt{r^2 - b^2} + \frac{M r}{\sqrt{r^2 - b^2}} - \frac{Q^2r \sqrt{1 - \frac{b^2}{r^2}}\text{Hypergeometric2F1}\left[ -\frac{1}{2}, 1,\frac{1}{2},  1 - \frac{r^2}{b^2} \right]}{2 (-b^2 + r^2)} \right) \right|_{r_0}^{r_D},
\end{aligned}
\end{equation}
where Hypergeometric2F1 is a hypergeometric function.\\
Note that the above result is for weak fields. As for strong fields, since Eq.\eqref{eq41} can not be directly integrated into a function with an explicit expression, numerical integration is directly adopted for calculation in the following section.
Furthermore, following the analytic-continuation prescription in Ref.~\cite{Alexandre:2025qip}, the imaginary lower-limit contributions cancel between the incoming and outgoing segments when $r_0<|b|$. We therefore retain only the real parts of the weak-field analytical phases. For the strong-field calculation, we use the unfactorized integrand in Eq.\eqref{eq13}, for which $1-A(r)b^2/r^2$ is nonnegative throughout the integration interval.

\subsubsection{Impact parameter $b$}
Substituting Eqs.\eqref{eq40} and \eqref{eq2} into Eq.\eqref{eq15} yields the relationship between $b$ and $r_{0}$ as follow
\begin{equation}\label{eq44}
r_0^2 = b^2 \left(1 - \frac{2M}{r_0} + \frac{Q^2}{r_0^2}\right).
\end{equation}
Hence, we have\\
\begin{equation}\label{eq:b_squared_expr}
b^2 = \frac{r_0^2}{1 - \frac{2M}{r_0} + \frac{Q^2}{r_0^2}}.
\end{equation}
Adopting the weak-field approximation, we obtain:
\begin{equation}\label{eq45}
\begin{split}
b = \frac{r_0}{\sqrt{1+ \frac{Q^2}{r_0^2}}} \times \left(1 + \frac{Mr_0}{Q^2 + r_0^2}\right) \quad \left( b>0, \, r_0>0 \right),\\
-b = \frac{r_0}{\sqrt{1+ \frac{Q^2}{r_0^2}}} \times \left(1 + \frac{Mr_0}{Q^2 + r_0^2}\right) \quad \left( b<0, \, r_0>0 \right).\\
\end{split}
\end{equation}

To express the impact parameter with the coordinates of the source and the detector in Fig.\ref{fig:lens}, we
employ the deflection angle under the Reissner--Nordstr\"om metric\cite{Sasaki:2023rdf}, i.e.,
\begin{equation}\label{eq46}
\delta = \frac{4M}{b} + \frac{3\pi}{4} \left(5 - \frac{Q^2}{M^2}\right) \left(\frac{M}{b}\right)^2 + \left(\frac{128}{3} - \frac{16Q^2}{M^2}\right) \left(\frac{M}{b}\right)^3.
\end{equation}
Note that for $b<0$, $b$ is replaced by $-b$ in the above formula.

According to Fig.\ref{fig:lens}, the following geometric relationship can be obtained:
\begin{equation}\label{eq25}
\delta \sim -\frac{y_D' - b}{x_D'},
\end{equation}
where $x' = x\cos\gamma + y\sin\gamma$, $y' = -x\sin\gamma + y\cos\gamma$, and $\sin\gamma = b/r_S$, with $\gamma$ being the rotation angle between the two coordinate systems.

For the weak-field case, taking the first-order term of $M$ in Eq.\eqref{eq46} and substituting it into Eq. \ref{eq25}, we obtain:
\begin{equation}\label{eq47}
-b - \frac{b \, x_D}{r_S} + \frac{4M y_D}{r_S} + \sqrt{1 - \frac{b^2}{r_S^2}} \left(\frac{4Mx_D}{b} + y_D \right) = 0.
\end{equation}
For the strong-field case, considering the second-order term of $M$ in Eq.\eqref{eq46}, we get
\begin{equation}\label{eq48}
-b-\frac{b x_D}{r_S}+\frac{b y_D}{r_S}\left(\frac{4M}{b}+\frac{3\pi(5M^2-Q^2)}{4b|b|}\right)+\sqrt{1-\frac{b^2}{r_S^2}}\left[\left(\frac{4M}{b}+\frac{3\pi(5M^2-Q^2)}{4b|b|}\right)x_D+y_D\right]=0.
\end{equation}
Solving the Eqs.\eqref{eq47} and Eq.\eqref{eq48} for $b$, the lensed neutrino oscillation probabilities can be obtained with the given location coordinates.
Furthermore, in the oscillation probability plots, we replace  $x_{D}$ and $y_{D}$ with the polar coordinates $r_{D}\cos\phi$ and $r_{D}\sin\phi$, respectively.

\subsection{Hayward metric}
\subsubsection{Oscillation Phase}
The line element of the Hayward metric is written as follows\cite{Hayward:2005gi}
\begin{equation}\label{eq16}
ds^{2} = -F(r)dt^{2} + \frac{dr^{2}}{F(r)} + r^{2}d\Omega^{2},F(r)=1-\frac{2M r^2}{r^3 + 2M l^2}.
\end{equation}
Substitute Eqs.\eqref{eq16} and \eqref{eq2} into Eq.\eqref{eq13} to obtain
\begin{equation}\label{eq17}
\begin{split}
\Phi_{k}\left(r_{S} \rightarrow r_{0} \rightarrow r_{D}\right) = \frac{m_{k}^{2}}{2E_{0}} \int_{r_{0}}^{r_{S}} \frac{1}{\sqrt{1 - \frac{b^{2}}{r^{2}}}} \left( 1 + \frac{2b^{2} M r^{2}}{(-b^{2} + r^{2})(2l^2M + r^{3})} \right)^{-1/2} dr\\
+ \frac{m_{k}^{2}}{2E_{0}} \int_{r_{0}}^{r_{D}} \frac{1}{\sqrt{1 - \frac{b^{2}}{r^{2}}}} \left( 1 + \frac{2b^{2} M r^{2}}{(-b^{2} + r^{2})(2l^2M + r^{3})} \right)^{-1/2} dr.
\end{split}
\end{equation}
Taking $\frac{2b^2 M r^2}{(-b^2 + r^2)(2l^2M + r^3)}$ as a small quantity,
Eq.\eqref{eq17} can be approximated as
\begin{equation}\label{eq18}
\begin{split}
\Phi_{k}\left(r_{S} \rightarrow r_{0} \rightarrow r_{D}\right) = \frac{m_{k}^{2}}{2E_{0}} \int_{r_{0}}^{r_{S}} \left( \frac{1}{\sqrt{1 - \frac{b^{2}}{r^{2}}}} - \frac{b^{2} M r^{2}}{\sqrt{1 - \frac{b^{2}}{r^{2}}}(-b^{2} + r^{2})(2l^2M + r^{3})} \right) dr\\
+\frac{m_{k}^{2}}{2E_{0}} \int_{r_{0}}^{r_{D}} \left( \frac{1}{\sqrt{1 - \frac{b^{2}}{r^{2}}}} - \frac{b^{2} M r^{2}}{\sqrt{1 - \frac{b^{2}}{r^{2}}}(-b^{2} + r^{2}) (2l^2M + r^{3})} \right)dr.
\end{split}
\end{equation}
Then considering the weak-field case where $\frac{M}{r}\ll1$, we obtain the result of the first-order approximation as follows
\begin{equation}\label{eq19}
\begin{split}
\Phi_{k}\left(r_{S} \rightarrow r_{0} \rightarrow r_{D}\right) = \frac{m_{k}^{2}}{2E_{0}} \int_{r_{0}}^{r_{S}} \left( \frac{1}{\sqrt{1 - \frac{b^{2}}{r^{2}}}} - \frac{b^{2} M}{r\sqrt{1 - \frac{b^{2}}{r^{2}}}(-b^{2} + r^{2})} + \frac{2b^2M^{2}l^{2}}{r^{4}\sqrt{1 - \frac{b^{2}}{r^{2}}}(-b^{2} + r^{2})} \right) dr\\
+\frac{m_{k}^{2}}{2E_{0}} \int_{r_{0}}^{r_{D}} \left( \frac{1}{\sqrt{1 - \frac{b^{2}}{r^{2}}}} - \frac{b^{2} M}{r\sqrt{1 - \frac{b^{2}}{r^{2}}}(-b^{2} + r^{2})} + \frac{2b^2M^{2}l^{2}}{r^{4}\sqrt{1 - \frac{b^{2}}{r^{2}}}(-b^{2} + r^{2})} \right) dr.
\end{split}
\end{equation}
After integrating Eq.\eqref{eq19}, we have
\begin{equation}
\begin{aligned}
\Phi_k &=\left. \frac{m_k^2}{2E_0} \left( \sqrt{r^2 - b^2} + \frac{M r}{\sqrt{r^2 - b^2}} - \frac{2l^2 M^2 \sqrt{1 - \frac{b^2}{r^2}} \ r\ Hypergeometric2F_1\left[- \frac{1}{2}, {2}, \frac{1}{2}, 1 - \frac{r^2}{b^2} \right]}{b^2(-b^2 + r^2)}\right)  \right\rvert_{r_0}^{r_S}\\
&\quad + \left. \frac{m_k^2}{2E_0} \left( \sqrt{r^2 - b^2} + \frac{M r}{\sqrt{r^2 - b^2}} - \frac{2l^2 M^2 \sqrt{1 - \frac{b^2}{r^2}} \ r\ Hypergeometric2F_1\left[- \frac{1}{2}, {2}, \frac{1}{2}, 1 - \frac{r^2}{b^2} \right]}{b^2(-b^2 + r^2)}\right)  \right\rvert_{r_0}^{r_D}.
\end{aligned}
\end{equation}
Here the imaginary contributions at the lower integration limit are treated using the same analytic-continuation prescription as in the RN case.
\subsubsection{Impact parameter $b$}
Substituting Eqs.\eqref{eq16} and \eqref{eq2} into Eq.\eqref{eq15}, we get the relationship between $b$ and $r_{0}$ as follows
\begin{equation}\label{eq21}
r_0^2 = b^2 \left( 1 - \frac{2M r_0^2}{r_0^3 + 2l^2M} \right).
\end{equation}
Thus, we have
\begin{equation}\label{eq22}
b^2 = \frac{r_0^2}{1 - \frac{2M r_0^2}{r_0^3 + 2l^2M}}.
\end{equation}
Adopting the weak-field approximation, we obtain
\begin{equation}\label{eq23}
\begin{split}
b = r_0 \times \left( 1 + \frac{M r_0^2}{r_0^3 + 2l^2M} \right) \quad \left( b>0, \, r_0>0 \right),\\
-b = r_0 \times \left( 1 + \frac{M r_0^2}{r_0^3 + 2l^2M} \right) \quad \left( b<0, \, r_0>0 \right).
\end{split}
\end{equation}

The deflection angle of the Hayward metric reads\cite{Chiba:2017nml}
\begin{equation}\label{eq24}
\delta = \frac{4M}{b} + \frac{15\pi}{4} \left( \frac{M}{b} \right)^2 + \frac{128}{3} \left( \frac{M}{b} \right)^3.
\end{equation}
For the weak-field deflection angle, taking the first order of $M$ in Eq.\eqref{eq24} and substituting it into Eq.\eqref{eq25}, we obtain
\begin{equation}\label{eq26}
-b - \frac{b \, x_D}{r_S} + \frac{4M y_D}{r_S} + \sqrt{1 - \frac{b^2}{r_S^2}} \left(\frac{4Mx_D}{b} + y_D \right) = 0.
\end{equation}
For the strong-field case, considering  the second-order term of $M$ , we get
\begin{equation}\label{eq27}
-b-\frac{b x_D}{r_S}+\frac{b y_D}{r_S}\left(\frac{4M}{b}+\frac{15\pi M^2}{4b|b|}\right)+\sqrt{1-\frac{b^2}{r_S^2}}\left[\left(\frac{4M}{b}+\frac{15\pi M^2}{4b|b|}\right)x_D+y_D\right]=0.
\end{equation}

\section{Numerical results }	
\label{sec:Nume}
According to the analytical expressions obtained in the previous sections, we show numerical results on CPVs. For illustration, we consider the channel $\nu_{e}(\nu^{c}_{e})\rightarrow\nu_{\mu}(\nu^{c}_{\mu})$.
The mixing parameters from NuFit-6.0\cite{Esteban:2024eli} are taken as $\Delta m_{21}^2 = 7.49 \times 10^{-5} \ \text{eV}^2,  \quad\Delta m_{31}^2 = 2.513 \times 10^{-3} \ \text{eV}^2  \left(\Delta m_{32}^2 = -2.484 \times 10^{-3}\text{eV}^2\right)$,
$\theta_{12} = 33.68^\circ(33.68^\circ), \quad\theta_{13} = 8.56^\circ(8.59^\circ), \quad\theta_{23} = 43.3^\circ \left( 47.9^\circ \right)$,
$\delta_{\text{CP}} = 212^\circ\left( 274^\circ \right)$ for normal (inverted) mass ordering. The neutrino energy is $E_0=10\text{MeV}$. The location parameters are chosen as
$r_D = 10^8\text{km}, \quad r_S = 10^5 r_D$, with the angle parameter of the detector $\phi$ varying in a tiny range. For the mass of the lens object, we set $M=M_{\odot}$ in the weak-field case, and $M=4*10^4M_{\odot}$  in the strong-field case
where the oscillation phases are calculated numerically without approximations and the second-order terms in the deflection angles are considered. 

\begin{figure}
		\centering
		\begin{subfigure}{\textwidth}
			\includegraphics[width=15cm,height=6cm]{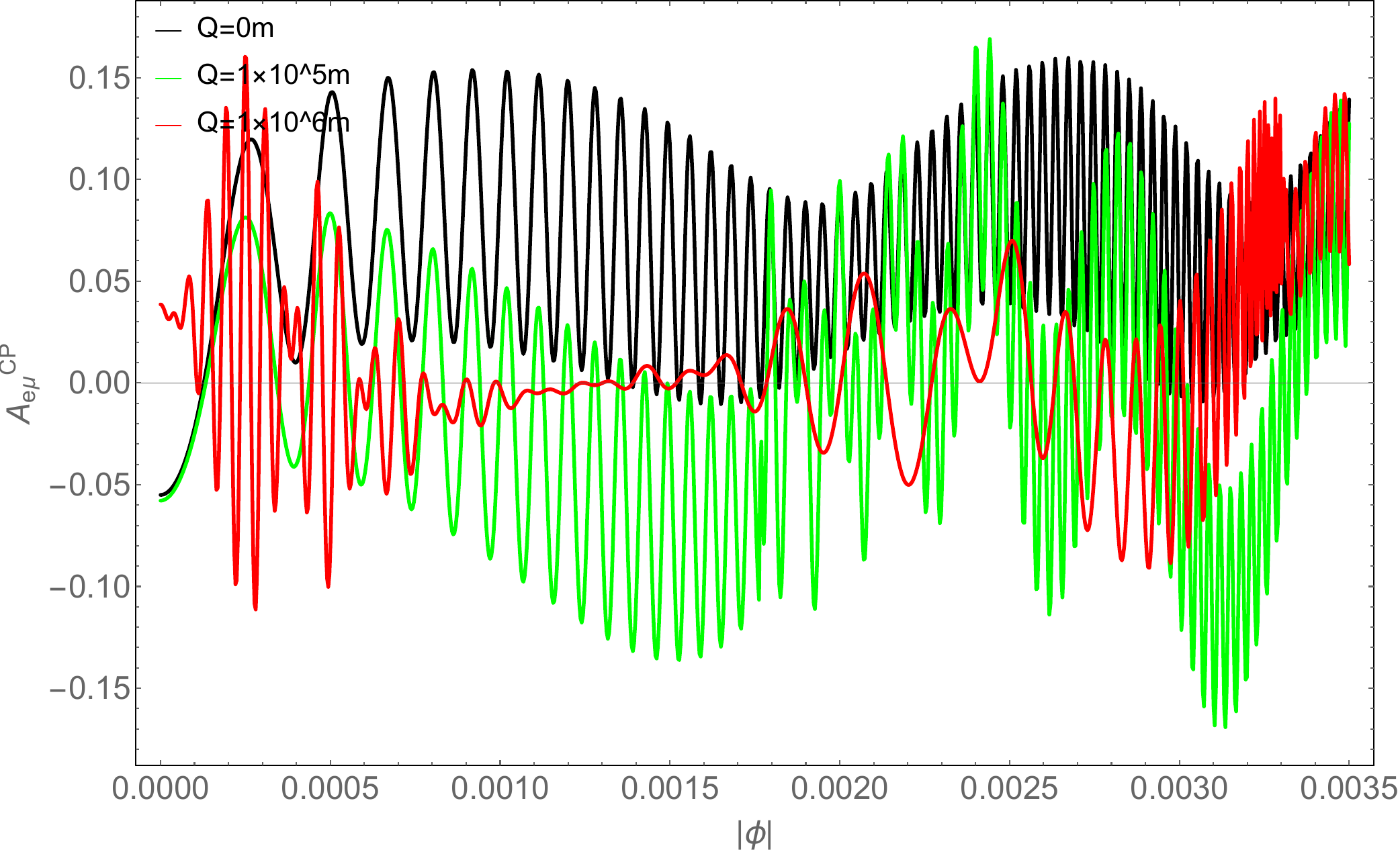}
			\caption{Normal ordering}
		\end{subfigure}\hfill

		\begin{subfigure}{\textwidth}
			\includegraphics[width=15cm,height=6cm]{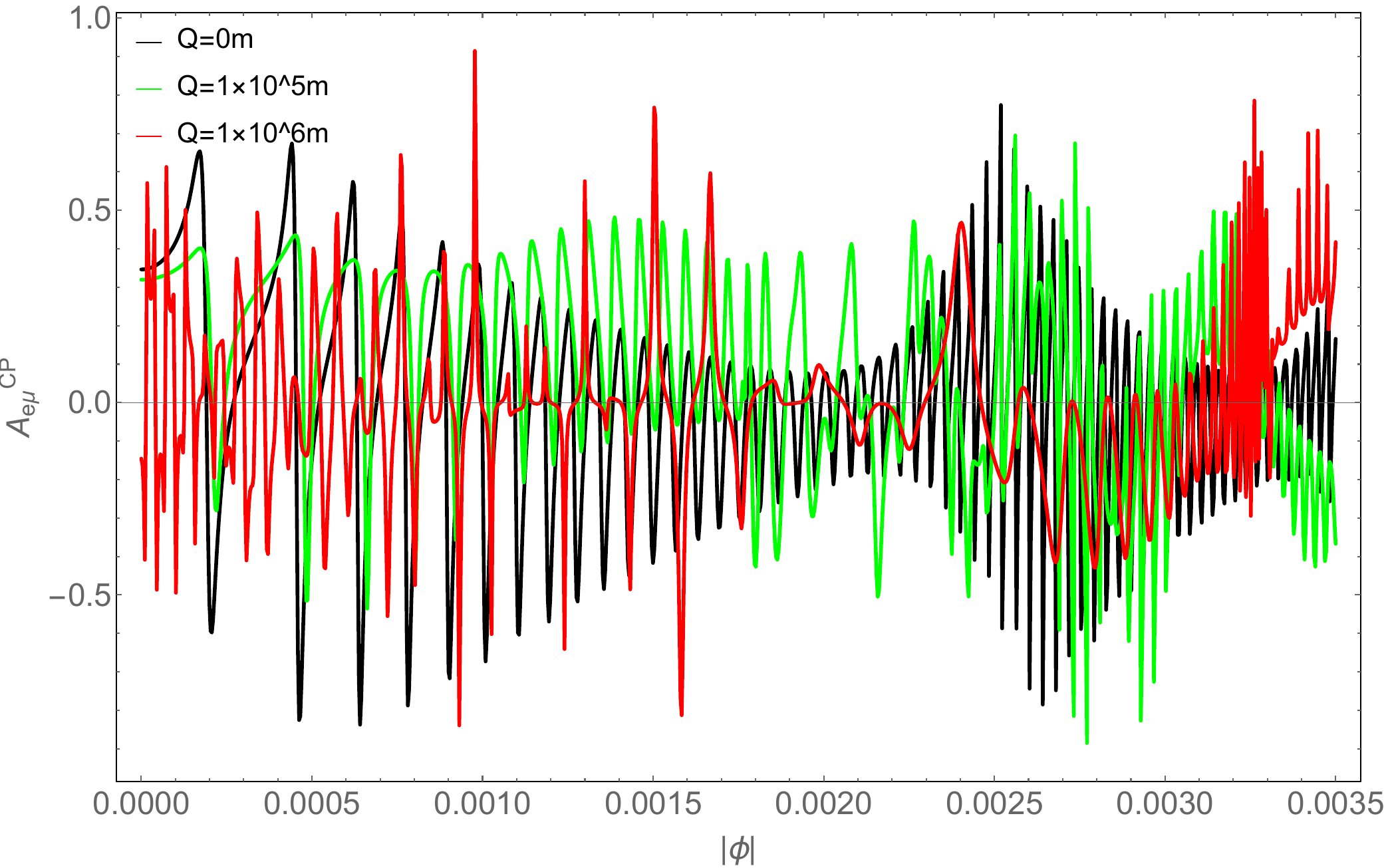}
			\caption{Inverted ordering}
		\end{subfigure}\hfill
	\caption{CP violation  including the lensing effects of RN metric  under weak-field, with $m_{l}=0$. }
		\label{fig:RNCPWea}
	\end{figure}

\begin{figure}
		\centering
		\begin{subfigure}{\textwidth}
			\includegraphics[width=15cm,height=6cm]{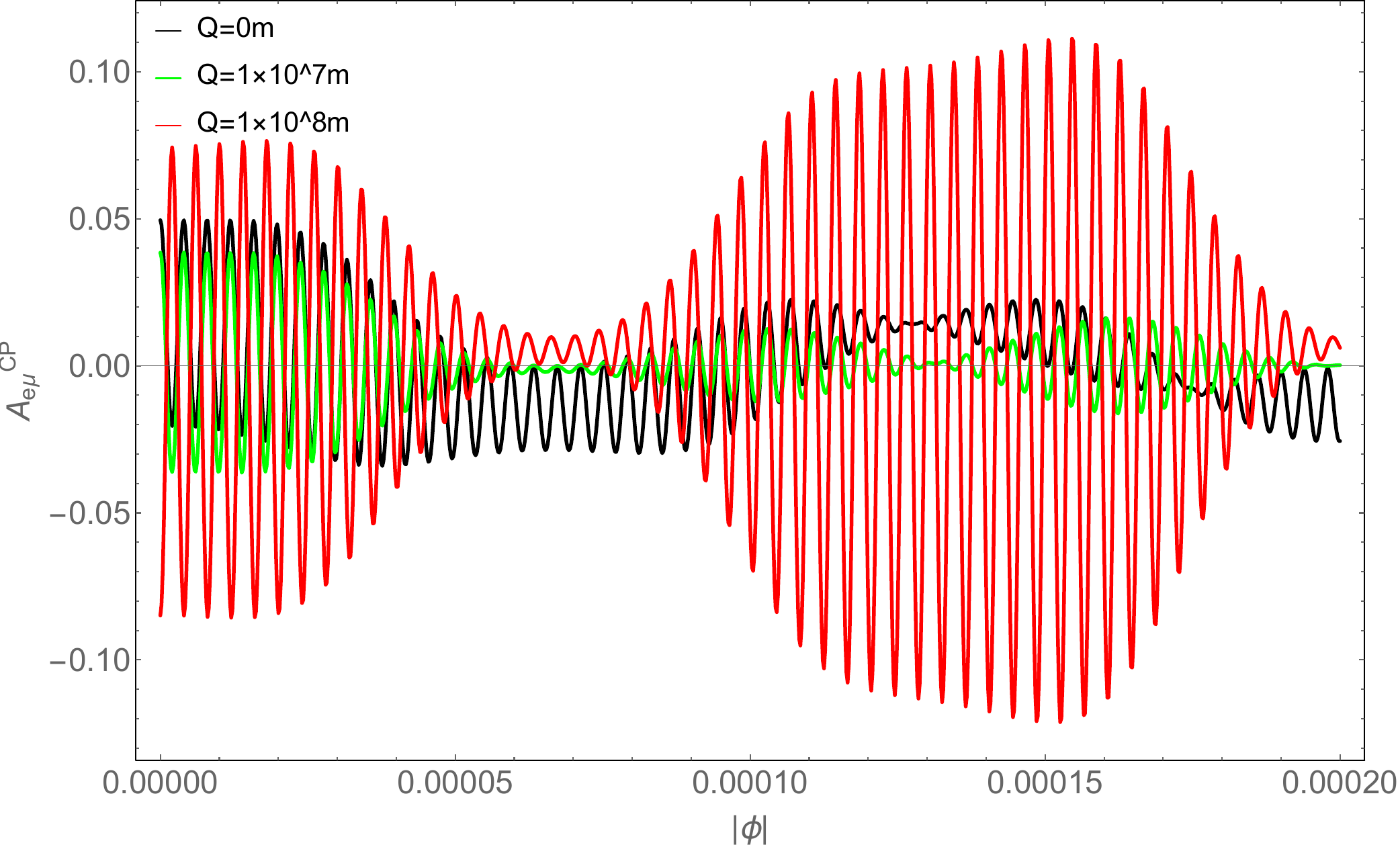}
			\caption{Normal ordering}
			\end{subfigure}\hfill

		\begin{subfigure}{\textwidth}
			\includegraphics[width=15cm,height=6cm]{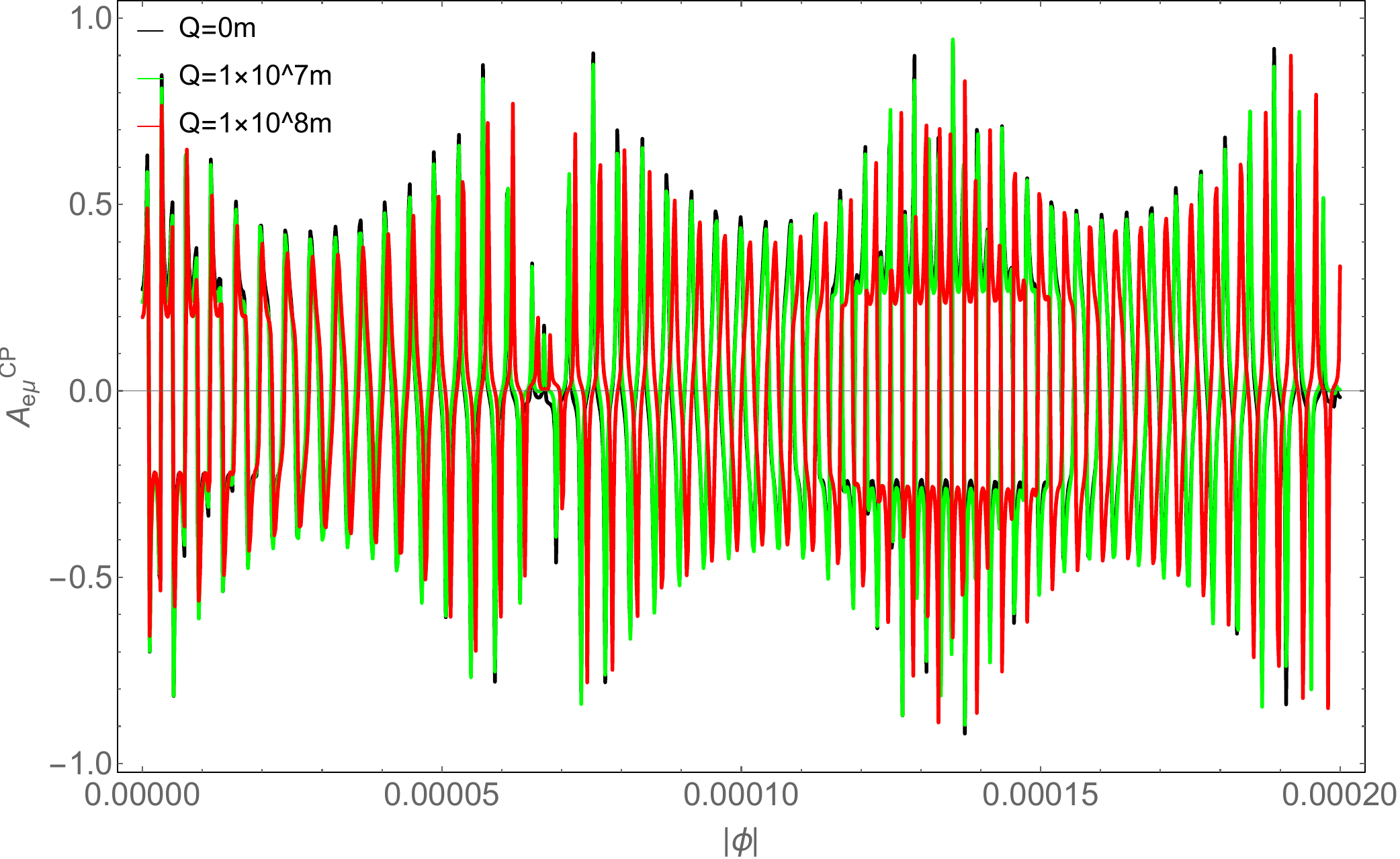}
			\caption{Inverted ordering}\end{subfigure}\hfill

		\caption{CP violation  including the lensing effects of  RN metric  under strong-field, with $m_{l}=0$. }
		\label{fig:RNCPStr}
	\end{figure}

\begin{figure}
		\centering
		\begin{subfigure}{\textwidth}
			\includegraphics[width=15cm,height=6cm]{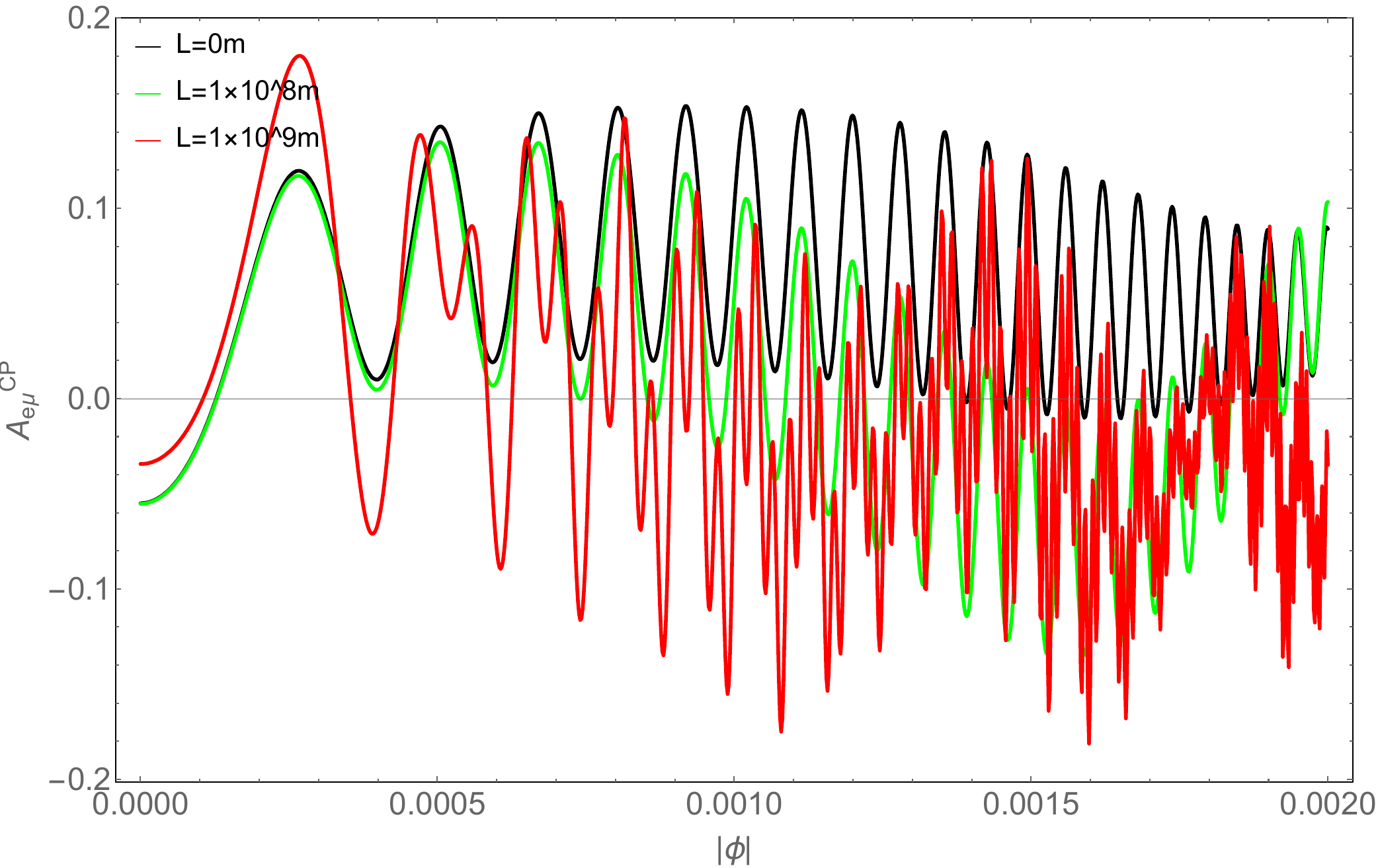}
			\caption{Normal ordering}
		\end{subfigure}\hfill

		\begin{subfigure}{\textwidth}
			\includegraphics[width=15cm,height=6cm]{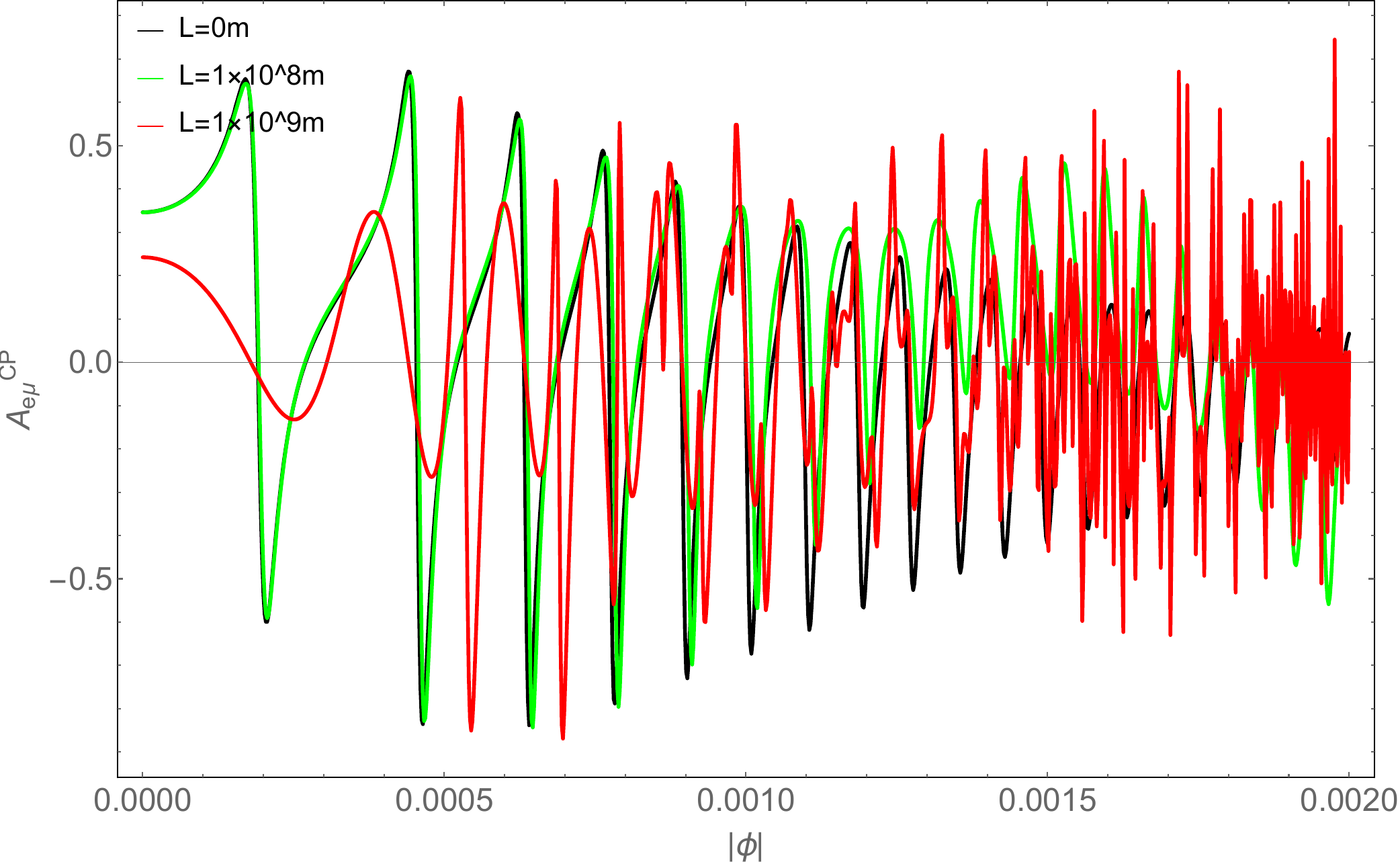}
			\caption{Inverted ordering}
		\end{subfigure}\hfill
	\caption{CP violation  including the lensing effects of HA metric under weak-field,  with $m_{l}=0$. }
		\label{fig:HACPWea}
	\end{figure}

\begin{figure}
		\centering
		\begin{subfigure}{\textwidth}
			\includegraphics[width=15cm,height=6cm]{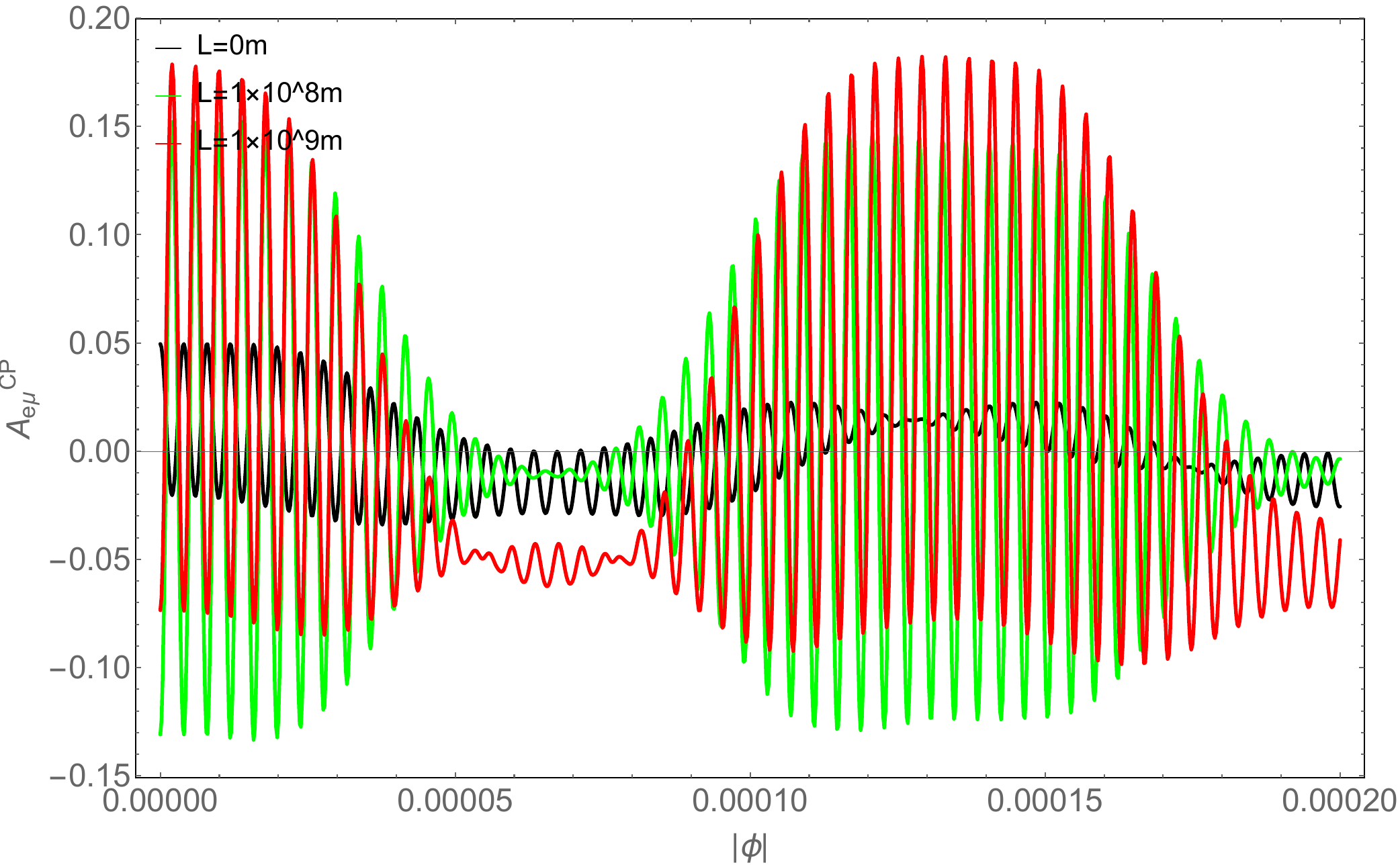}
			\caption{Normal ordering}
			\end{subfigure}\hfill

		\begin{subfigure}{\textwidth}
			\includegraphics[width=15cm,height=6cm]{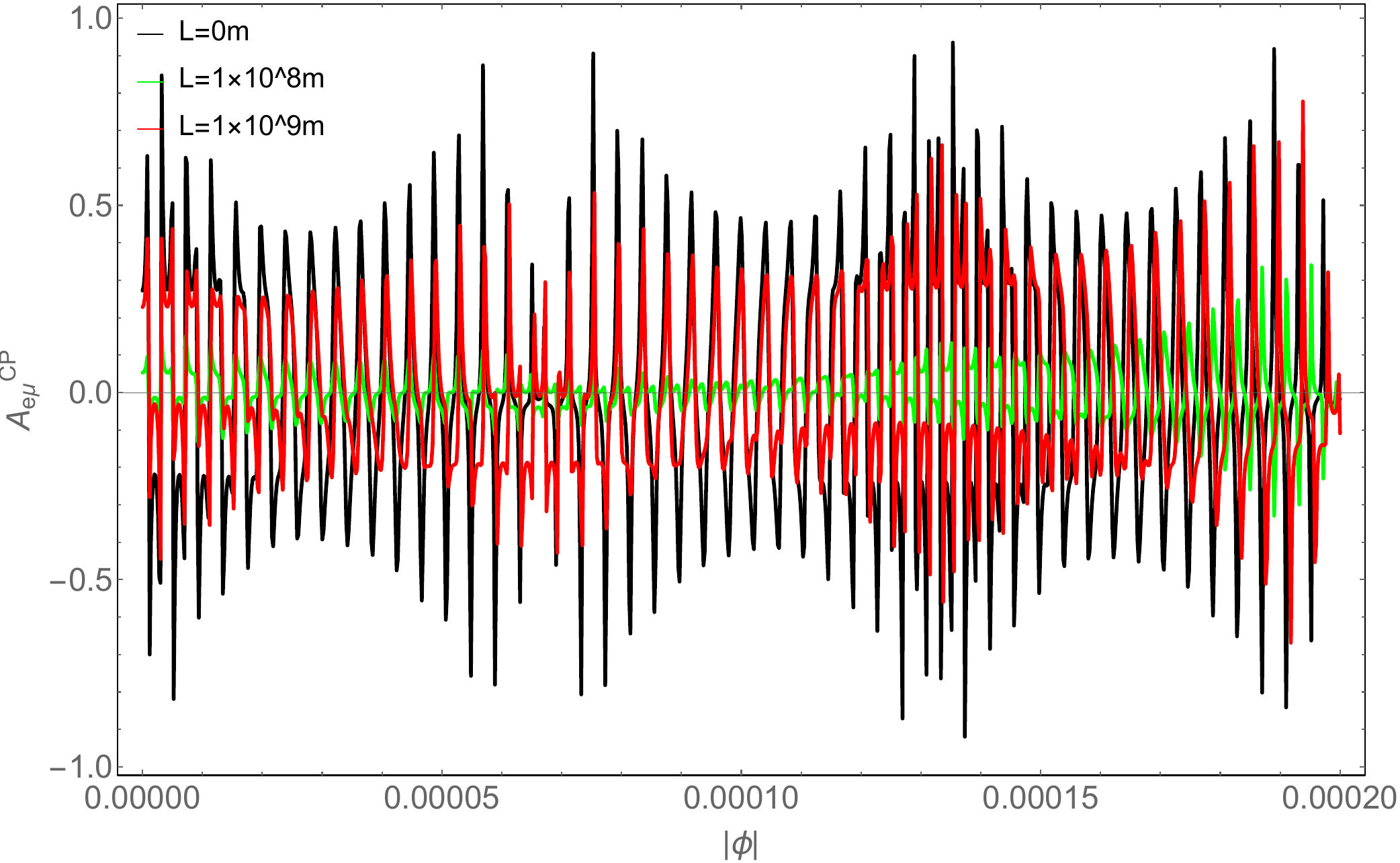}
			\caption{Inverted ordering}\end{subfigure}\hfill

		\caption{CP violation  including the lensing effects of HA metric under strong-field, with $m_{l}=0$. }
		\label{fig:HACPStr}
	\end{figure}
	
We first give the results with the lightest neutrino-mass $m_{l}=0 $.
Fig.~\ref{fig:RNCPWea} and Fig.~\ref{fig:RNCPStr} show the CPV including lensing effects of RN metric in the weak-field and strong-field case, respectively.
It is pronounced that the oscillation pattern is sensitive to the mass ordering. The oscillation amplitude with the inverted ordering (IO) is larger than that with the normal ordering (NO) both in the weak-field and strong-field case.
 This observation is partially due to the fact that $|\sin\delta^{IO}_{CP}|\sim2|\sin\delta^{NO}_{CP}|$.

 In the weak-field case, the oscillations of $A^{CP}_{e\mu}$ become more closely spaced with increasing $\phi$, while the charge parameter $Q$ significantly modifies the oscillation profile. In the strong-field case, the influence of $Q$ becomes more pronounced for NO. In particular, for $Q=1\times10^8\,\mathrm{m}$, the oscillation amplitude is significantly enlarged compared with the cases of $Q=0$ and $Q=1\times10^7\,\mathrm{m}$.

With respect to the lensing effects of non-singular gravity, Fig.~\ref{fig:HACPWea} - Fig.~\ref{fig:HACPStr} for HA metric show that the observations on the amplitudes
and periods still hold in this case. Furthermore, in the strong-field case, the metric parameter $l$ has a significant modulation effect on the amplitude of the CPV curve for both NO and IO.
For NO, the Schwarzschild metric with $l=0\,\mathrm{m}$ gives a smaller CP-violation amplitude than that in the HA metric. By contrast, for IO, the opposite behavior is observed for small values of $\phi$.

\begin{figure}
		\centering
		\begin{subfigure}{\textwidth}
			\includegraphics[width=15cm,height=6cm]{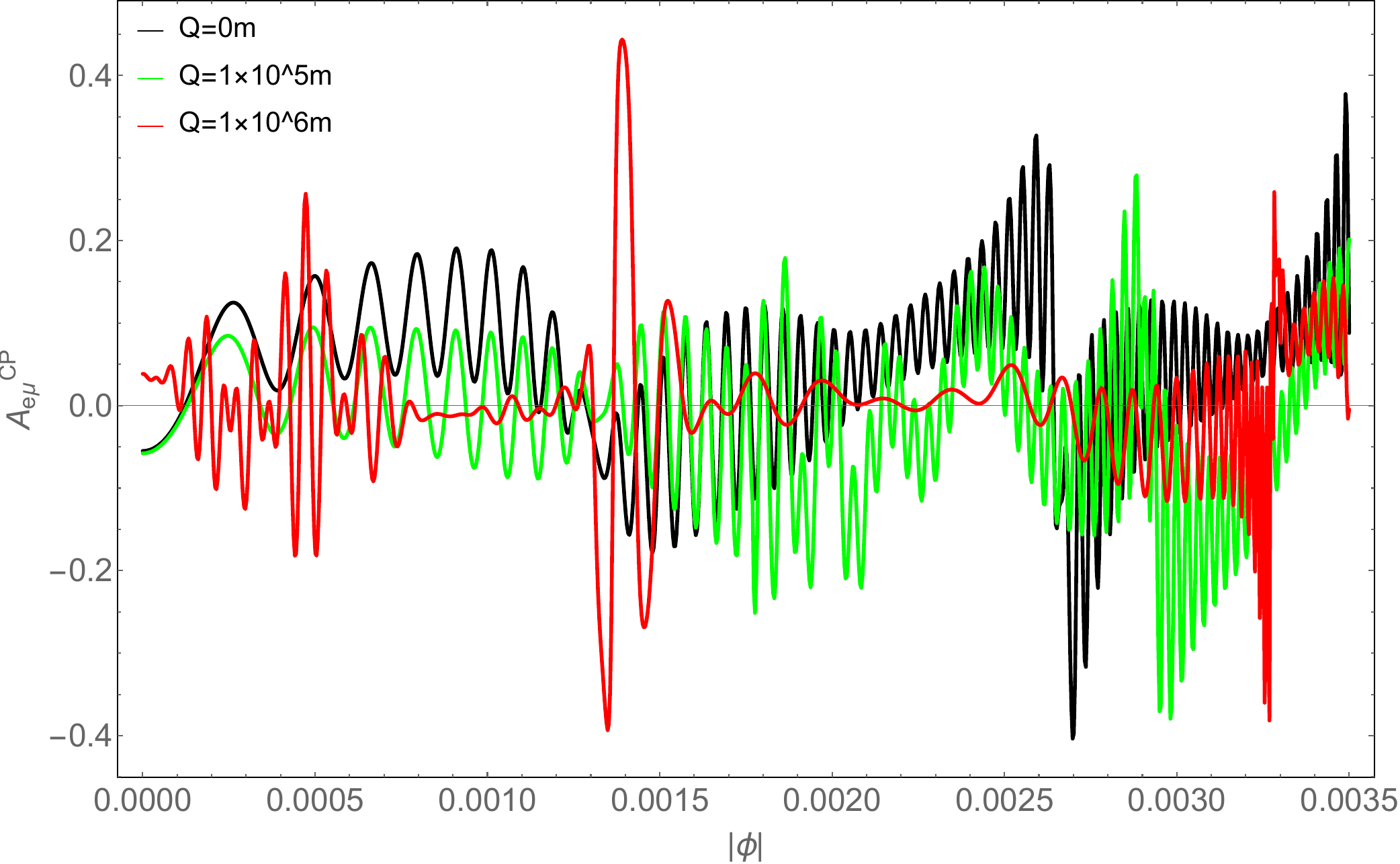}
			\caption{Normal ordering}
		\end{subfigure}\hfill

		\begin{subfigure}{\textwidth}
			\includegraphics[width=15cm,height=6cm]{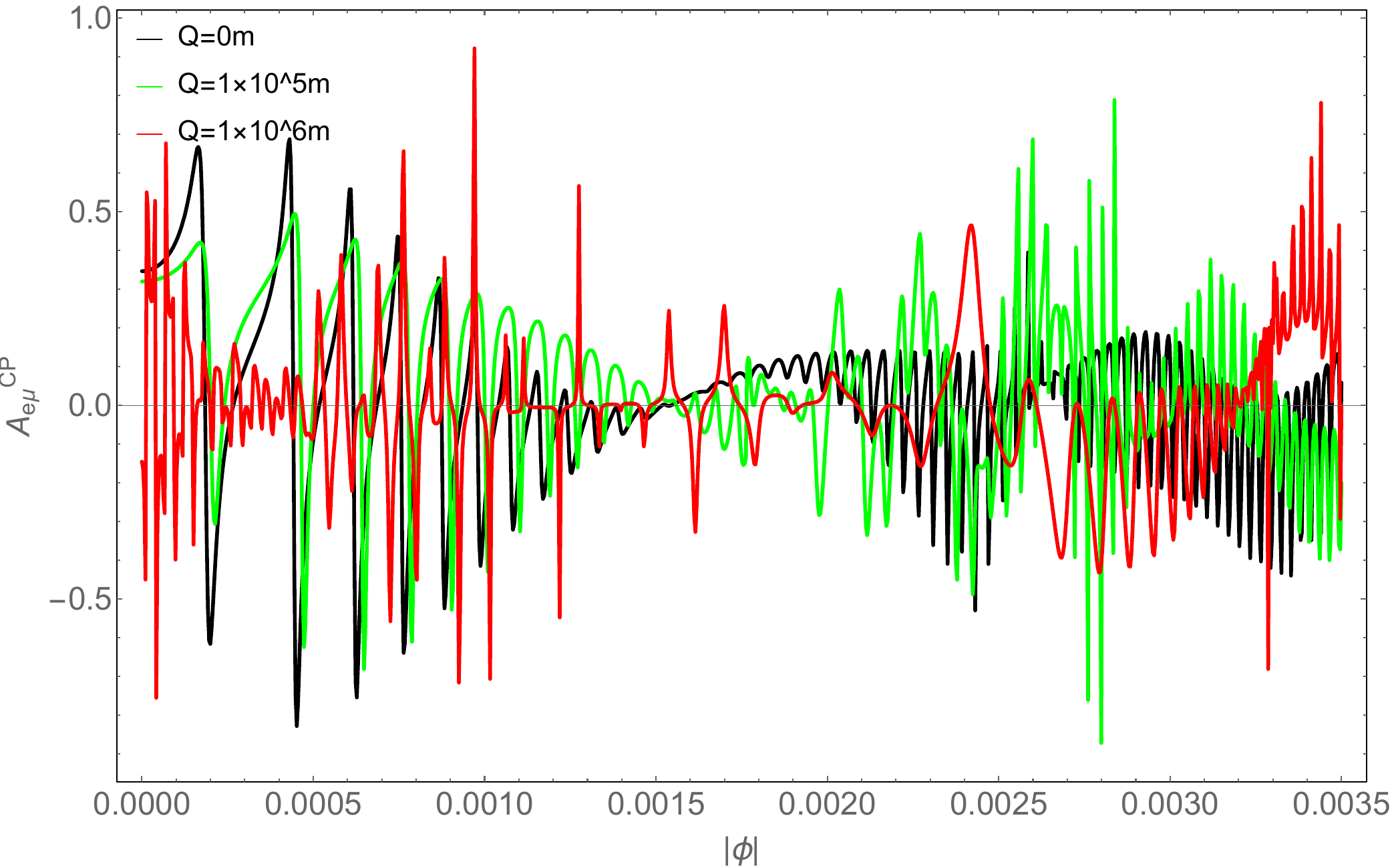}
			\caption{Inverted ordering}
		\end{subfigure}\hfill
	\caption{CP violation  including the lensing effects of RN metric  under weak-field, with $m_{l}=0.01 \text{eV}$. }
		\label{fig:RNCPWealm}
	\end{figure}

\begin{figure}
		\centering
		\begin{subfigure}{\textwidth}
			\includegraphics[width=15cm,height=6cm]{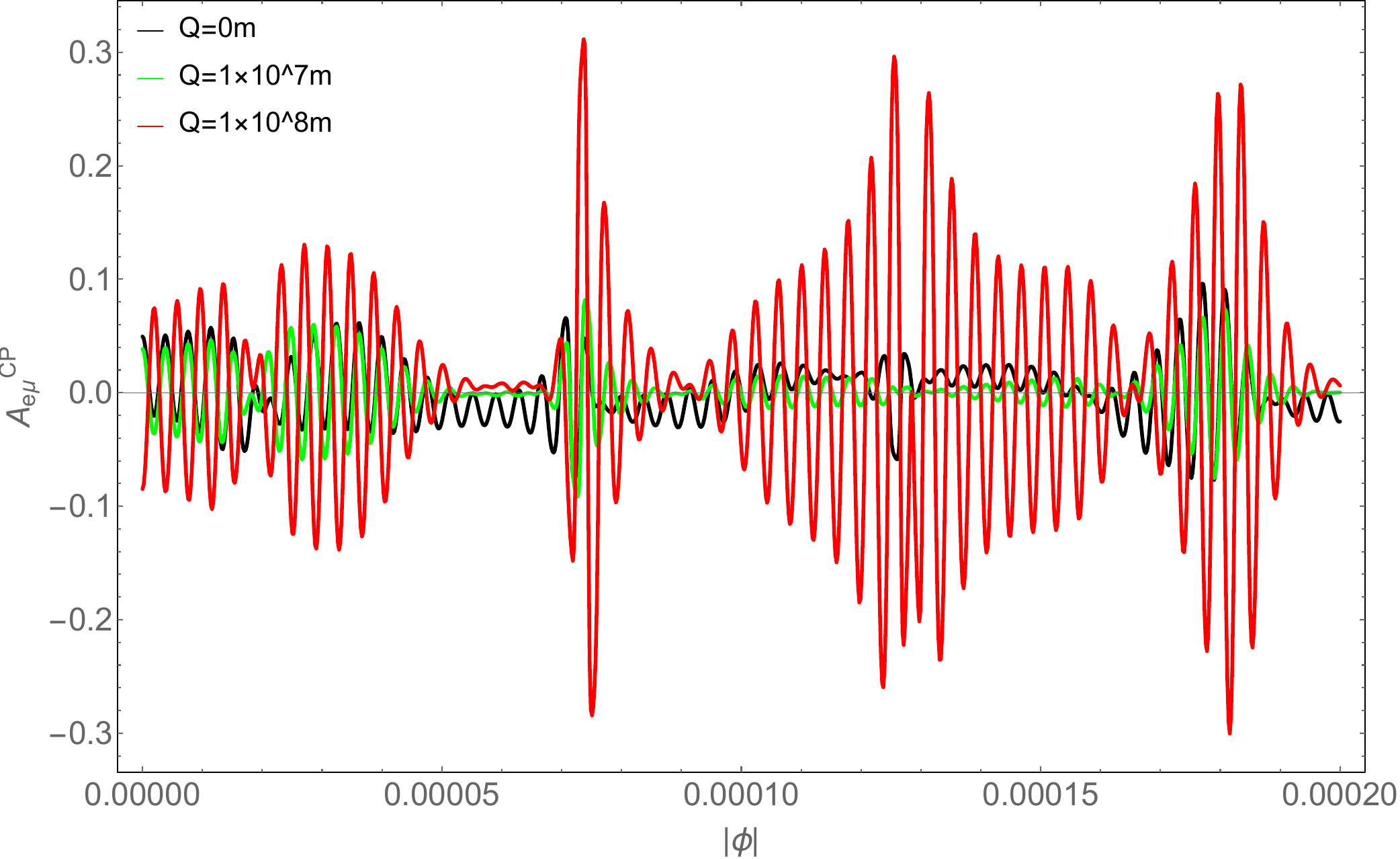}
			\caption{Normal ordering}
			\end{subfigure}\hfill

		\begin{subfigure}{\textwidth}
			\includegraphics[width=15cm,height=6cm]{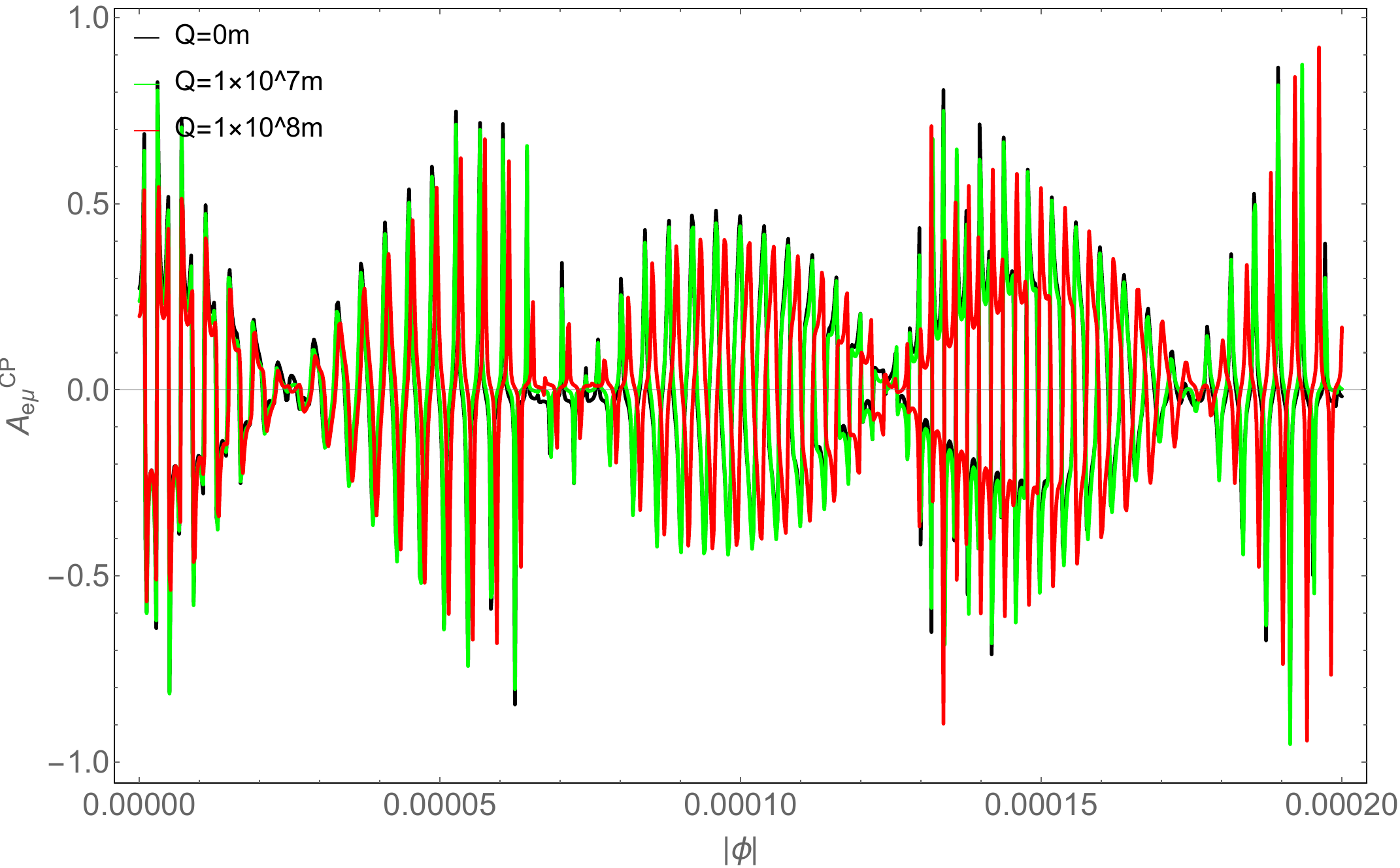}
			\caption{Inverted ordering}\end{subfigure}\hfill

		\caption{CP violation  including the lensing effects of  RN metric  under strong-field, with $m_{l}=0.01 \text{eV}$. }
		\label{fig:RNCPStrlm}
	\end{figure}

\begin{figure}
		\centering
		\begin{subfigure}{\textwidth}
			\includegraphics[width=15cm,height=6cm]{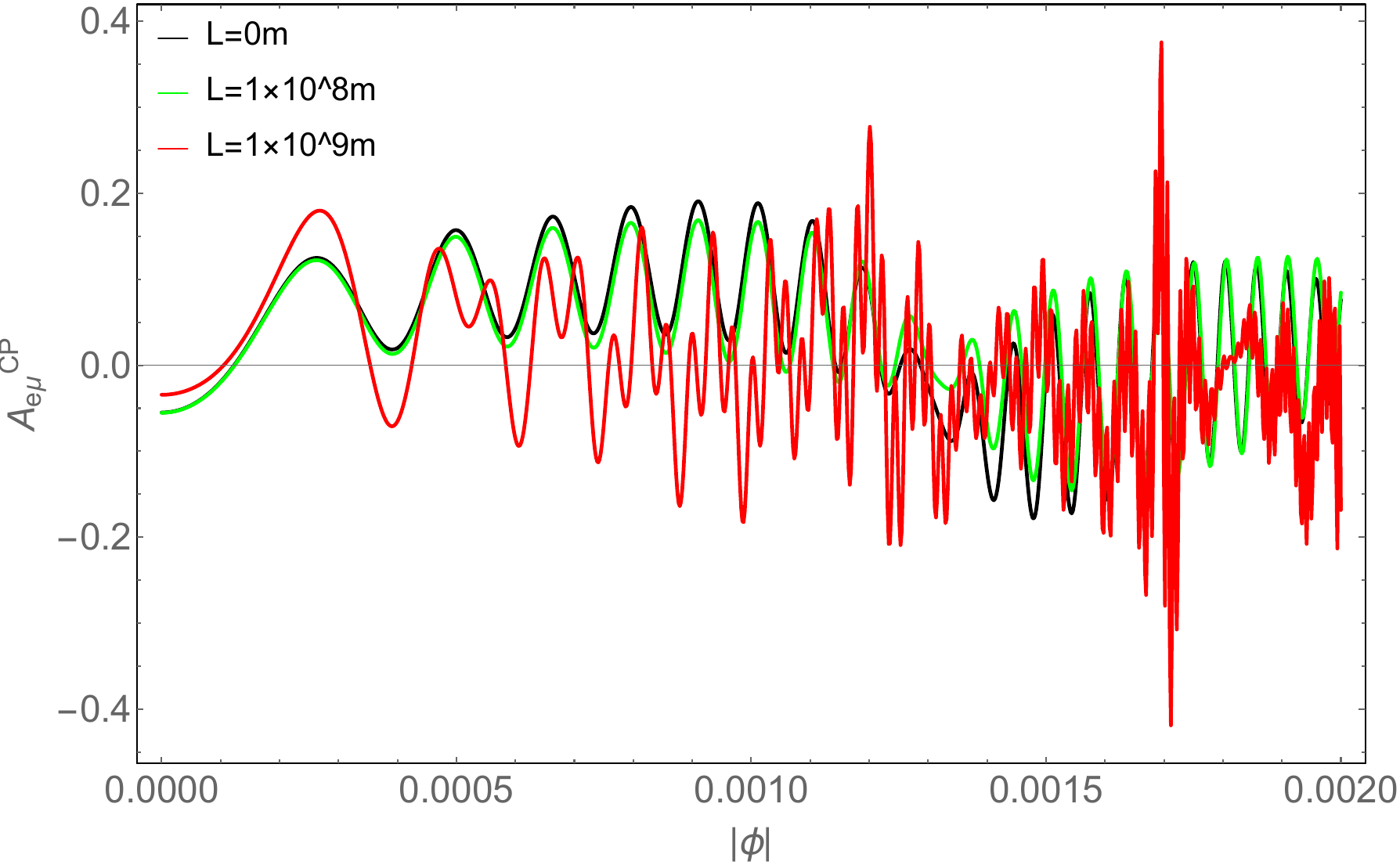}
			\caption{Normal ordering}
		\end{subfigure}\hfill

		\begin{subfigure}{\textwidth}
			\includegraphics[width=15cm,height=6cm]{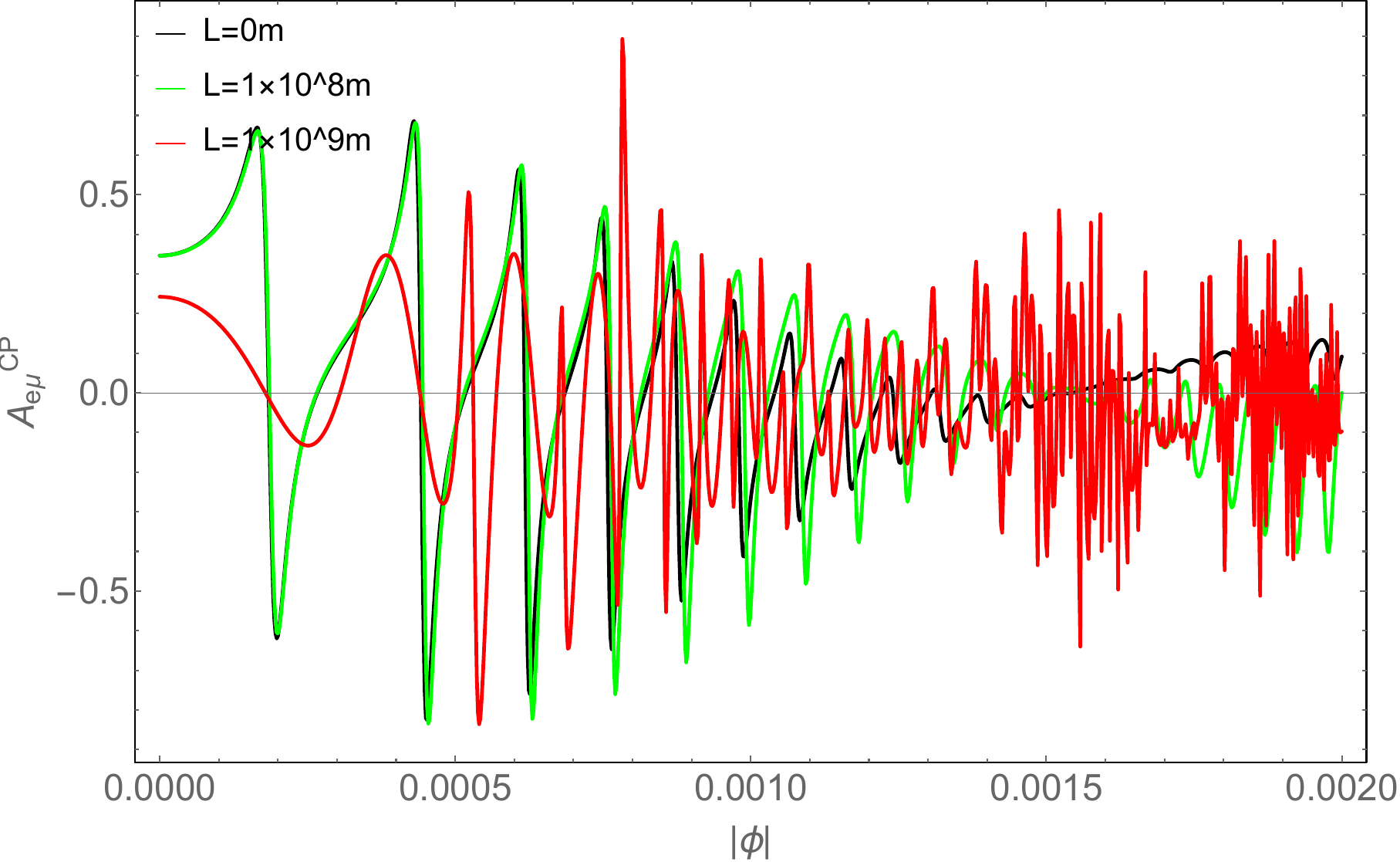}
			\caption{Inverted ordering}
		\end{subfigure}\hfill
	\caption{CP violation  including the lensing effects of HA metric under weak-field,  with $m_{l}=0.01 \text{eV}$. }
		\label{fig:HACPWealm}
	\end{figure}

\begin{figure}
		\centering
		\begin{subfigure}{\textwidth}
			\includegraphics[width=15cm,height=6cm]{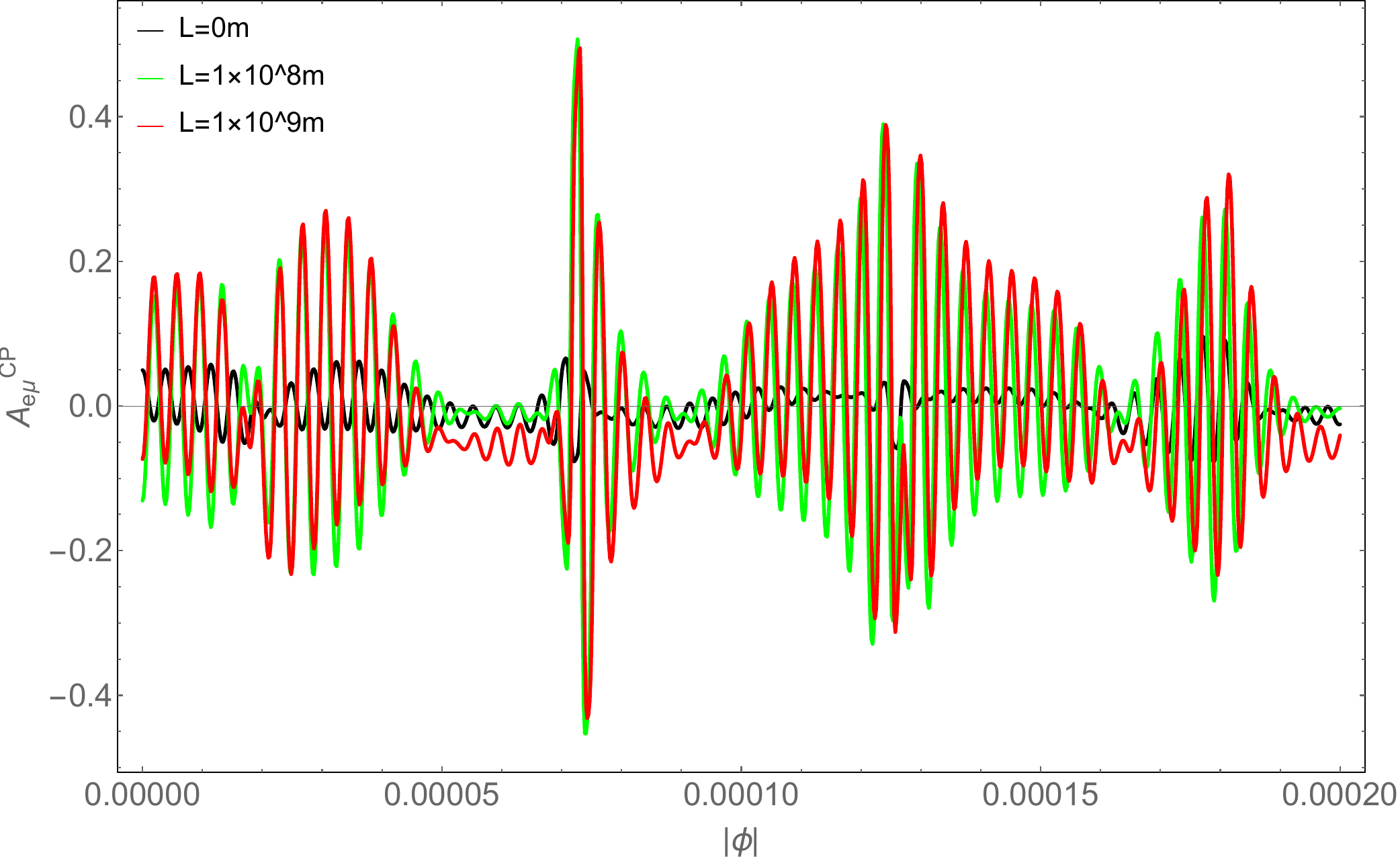}
			\caption{Normal ordering}
			\end{subfigure}\hfill

		\begin{subfigure}{\textwidth}
			\includegraphics[width=15cm,height=6cm]{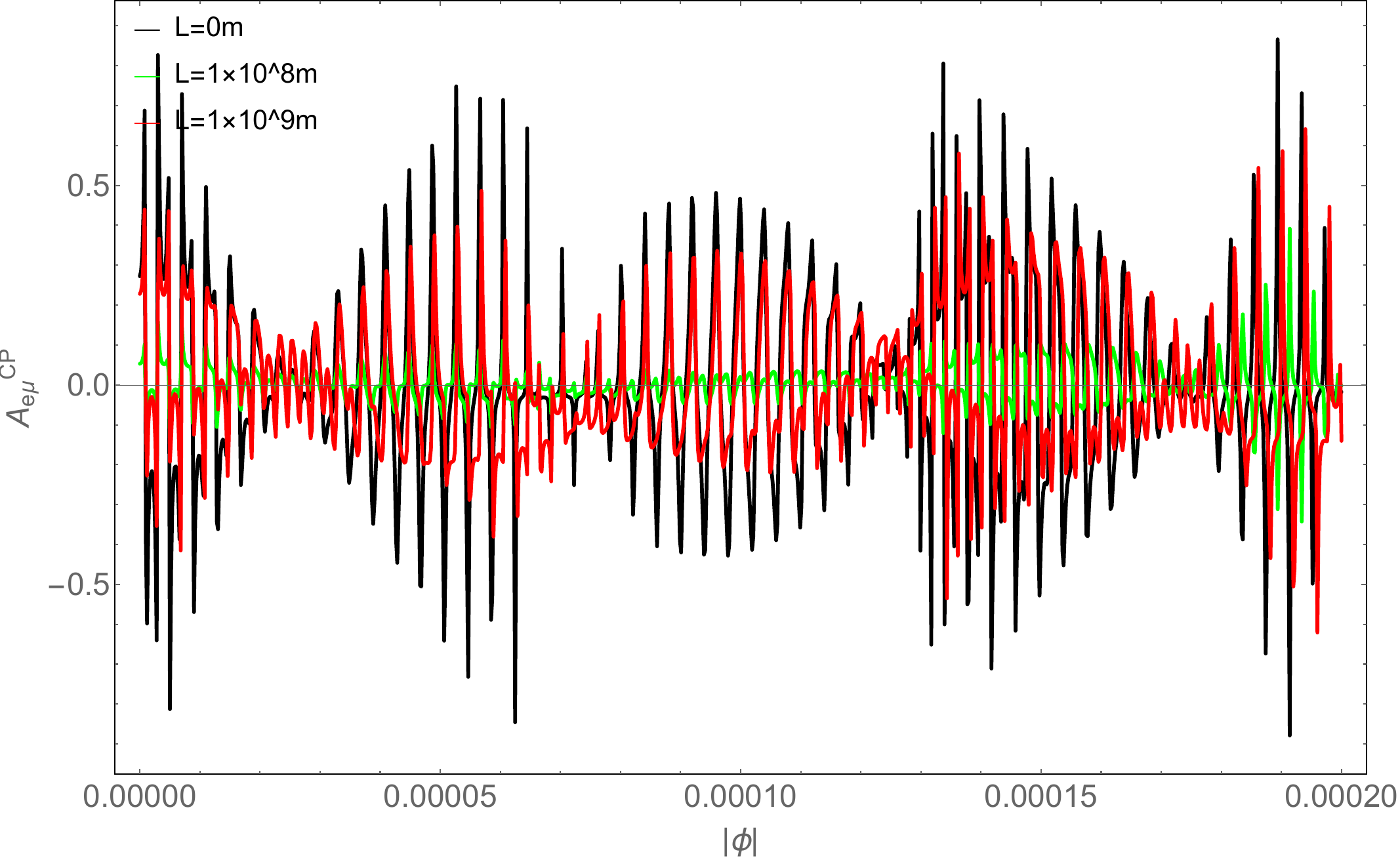}
			\caption{Inverted ordering}\end{subfigure}\hfill

		\caption{CP violation  including the lensing effects of HA metric under strong-field, with $m_{l}=0.01 \text{eV}$. }
		\label{fig:HACPStrlm}
	\end{figure}

As is known, the oscillation probabilities are in general dependent on the absolute masses of neutrinos in curved space-time. We now examine the consequences of the nonzero lightest mass with $m_{l}=0.01 \text{eV}$.
The results are shown in Fig.~\ref{fig:RNCPWealm} - Fig.~\ref{fig:HACPStrlm}.

For RN metric, in both the weak-field and strong-field regimes, the NO case exhibits a noticeably larger oscillation amplitude for $m_l=0.01\,\mathrm{eV}$. Furthermore, across all considered cases, increasing the lightest neutrino mass significantly modifies the oscillation profile of the CPV.
For HA metric, the effect of $m_l$ on the oscillation amplitude in the NO case is consistent with that observed for the RN metric. In addition, in the weak-field case with IO, the oscillation patterns corresponding to the two lightest neutrino masses are highly similar, see  Fig.~\ref{fig:HACPWea} and  Fig.~\ref{fig:HACPWealm}. Furthermore, for NO in the strong-field case with nonzero $m_l$, the difference between the oscillation curves for $l=1\times10^8\,\mathrm{m}$ and $l=1\times10^9\,\mathrm{m}$ is rather limited, see Fig.~\ref{fig:HACPStrlm}.
For a more complete comparison, the results for
$m_l=0.03\,\mathrm{eV}$ are presented in
Appendix~\ref{app:mass003} (see Figs.~\ref{fig:RNCPWealm003}--\ref{fig:HACPStrlm003}), covering both mass orderings
and both field-strength regimes for the RN and HA metrics.
Compared with the $m_l=0$ case, these results show qualitative trends similar to those found for $m_l=0.01\,\mathrm{eV}$.

\section{Conclusion }	
\label{sec:Conc}
In this work, the CPVs including GL effects in the case of 3-flavor neutrino oscillations were investigated systematically. The impacts on the oscillations from mass ordering, the absolute mass of neutrino, and the field-strength were considered with the RN, and HA metric. The analytical expressions for oscillation probabilities were derived under the weak-field approximation and the numerical method to calculate the results for strong fields were presented.

Assuming a vanishing lightest neutrino mass ($m_l = 0$), we find that the inverted ordering (IO) universally yields larger CPV oscillation amplitudes than the normal ordering (NO), establishing mass ordering as the primary observable baseline. The gravitational field then imposes two-tiered modulations: in the weak-field regime, increasing $\phi$ compresses the CPV curves, while in the strong-field regime, metric-specific parameters drastically reshape the signal---most notably, the RN charge exhibits a pronounced amplitude-amplification at $Q = 1 \times 10^{8}\,\mathrm{m}$, whereas the HA Schwarzschild limit ($l = 0\,\mathrm{m}$) produces an opposite trend between NO and IO. 

For nonzero $m_l$, we find that its sensitivity is geometry-dependent: the singular RN spacetime substantially reshapes all profiles, yet the regular HA spacetime nearly suppresses this sensitivity in the weak-field  case with IO. These results demonstrate that CPV signals encode both geometry and mass ordering, and that the imprints of $m_l$ are intrinsically correlated with the singular versus nonsingular structure of the background metric.
In particular, our observations show that the CPV observable can  circumvent the limitation that, in asymptotically flat
spacetime, gravity neither damps nor amplifies neutrino-oscillation amplitudes\cite{Wu:2026cfj,Li:2026olr}.

	\acknowledgments
This work was supported by the National Natural Science Foundation of China under Grant No. 12565015 and the Natural Science Foundation of Guangxi under Grant No. 2026GXNSFAA00640923.
	.
\clearpage
\appendix
\section{Results for $m_l=0.03\,\mathrm{eV}$}
\label{app:mass003}
\setcounter{figure}{0}
\renewcommand{\thefigure}{A\arabic{figure}}
\ifdefined\theHfigure
    \renewcommand{\theHfigure}{A.\arabic{figure}}
\fi
\begin{figure}[!htbp]
    \centering
    \begin{subfigure}{\textwidth}
        \includegraphics[width=15cm,height=6cm]{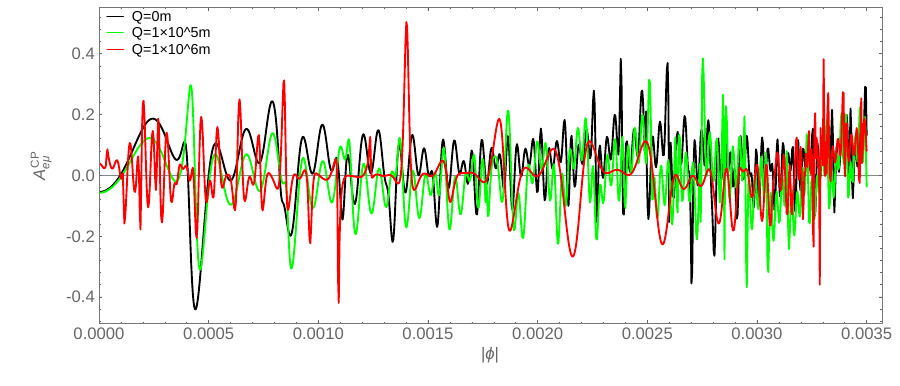}
        \caption{Normal ordering}
    \end{subfigure}\hfill

    \begin{subfigure}{\textwidth}
        \includegraphics[width=15cm,height=6cm]{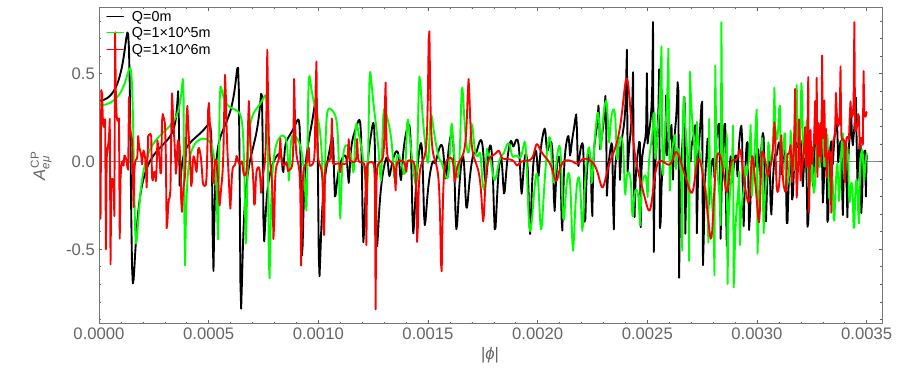}
        \caption{Inverted ordering}
    \end{subfigure}\hfill

    \caption{CP violation including the lensing effects of RN metric under weak-field, with $m_l=0.03\,\text{eV}$.}
    \label{fig:RNCPWealm003}
\end{figure}

\begin{figure}[!htbp]
    \centering
    \begin{subfigure}{\textwidth}
        \includegraphics[width=15cm,height=6cm]{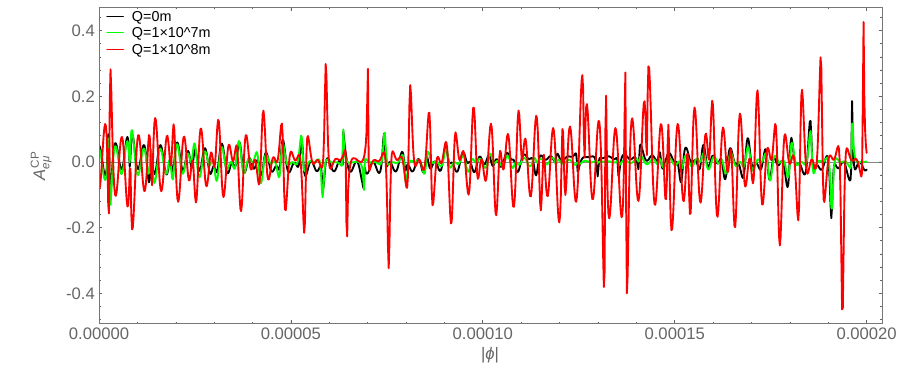}
        \caption{Normal ordering}
    \end{subfigure}\hfill

    \begin{subfigure}{\textwidth}
        \includegraphics[width=15cm,height=6cm]{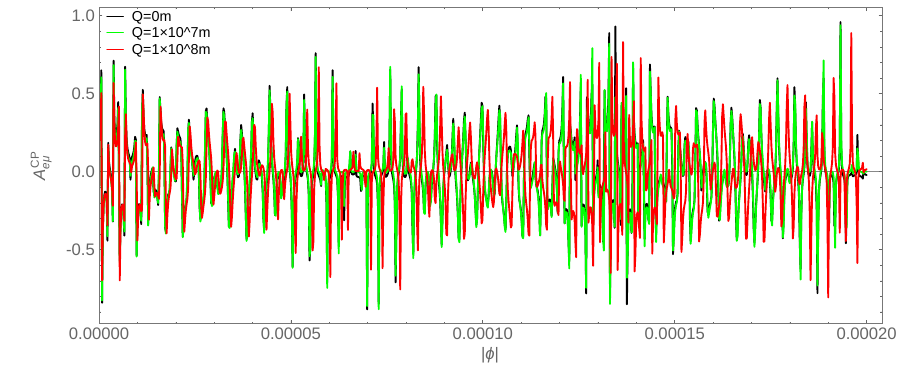}
        \caption{Inverted ordering}
    \end{subfigure}\hfill

    \caption{CP violation including the lensing effects of RN metric under strong-field, with $m_l=0.03\,\text{eV}$.}
    \label{fig:RNCPStrlm003}
\end{figure}

\begin{figure}[!htbp]
    \centering
    \begin{subfigure}{\textwidth}
        \includegraphics[width=15cm,height=6cm]{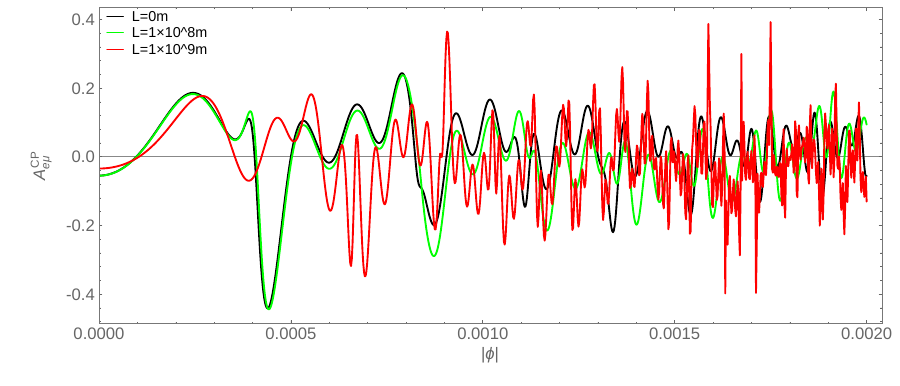}
        \caption{Normal ordering}
    \end{subfigure}\hfill

    \begin{subfigure}{\textwidth}
        \includegraphics[width=15cm,height=6cm]{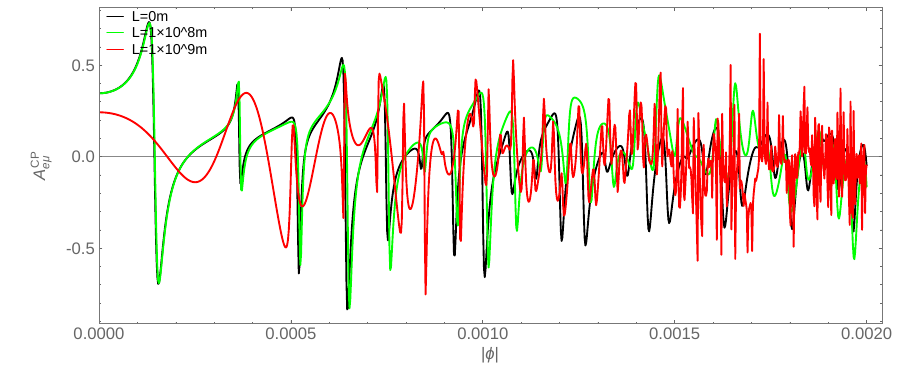}
        \caption{Inverted ordering}
    \end{subfigure}\hfill

    \caption{CP violation including the lensing effects of HA metric under weak-field, with $m_l=0.03\,\text{eV}$.}
    \label{fig:HACPWealm003}
\end{figure}

\begin{figure}[!htbp]
    \centering
    \begin{subfigure}{\textwidth}
        \includegraphics[width=15cm,height=6cm]{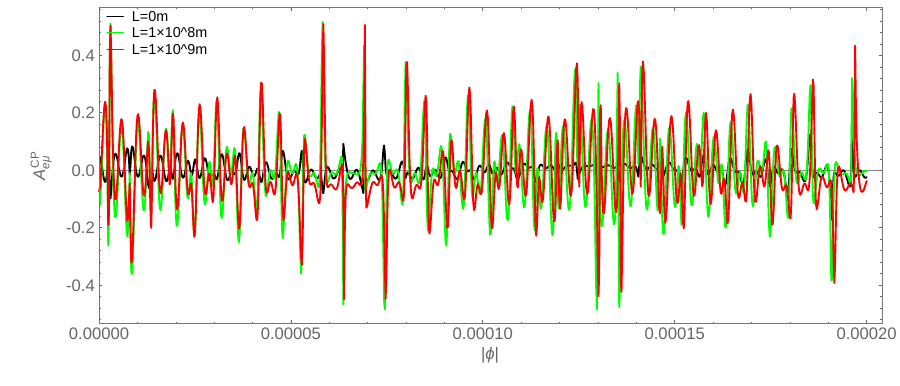}
        \caption{Normal ordering}
    \end{subfigure}\hfill

    \begin{subfigure}{\textwidth}
        \includegraphics[width=15cm,height=6cm]{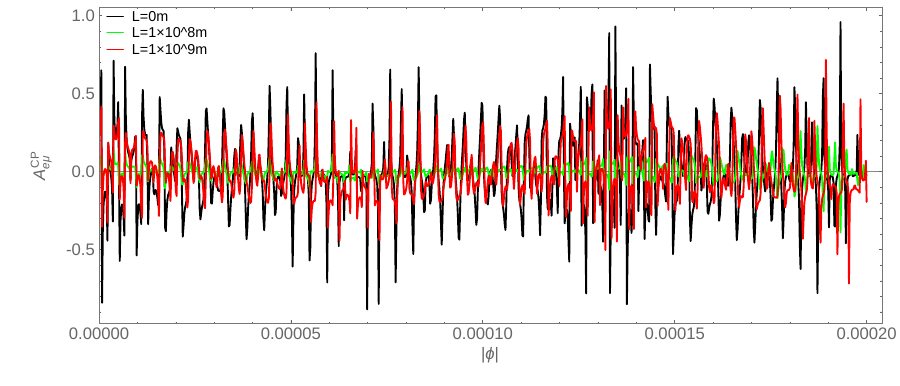}
        \caption{Inverted ordering}
    \end{subfigure}\hfill

    \caption{CP violation including the lensing effects of HA metric under strong-field, with $m_l=0.03\,\text{eV}$.}
    \label{fig:HACPStrlm003}
\end{figure}
\clearpage
	\bibliography{biblio}

\end{document}